\documentclass[12pt]{article}
\usepackage{graphicx}
\usepackage{aasms4}

\def\lapprox{\hbox{\lower .8ex\hbox{$\,\buildrel < \over\sim\,$}}}
\def\gapprox{\hbox{\lower .8ex\hbox{$\,\buildrel > \over\sim\,$}}}

\begin{document}

\title{New approaches to SNe Ia progenitors}

\author{Pilar Ruiz--Lapuente$^{1, 2}$}

\altaffiltext{1}{Instituto de F\'{\i}sica Fundamental, Consejo Superior de 
Investigaciones Cient\'{\i}ficas, c/. Serrano 121, E--28006, Madrid, Spain}
\altaffiltext{2}{Institut de Ci\`encies del Cosmos (UB--IEEC), c/. Mart\'{\i} i 
Franqu\'es 1, E--08028, Barcelona, Spain}

\begin{abstract}
Although Type Ia supernovae (SNe Ia) are a major tool in cosmology and  
play a key role in the chemical evolution of galaxies, the nature of
their progenitor systems (apart from the fact that they must content at least
one white dwarf, that explodes) remains largely unknown. In the last 
decade, considerable efforts have been made, both observationally and 
theoretically, to solve this problem. Observations have, however, 
revealed a previously ususpected variety of events, ranging from very 
underluminous outbursts to clearly overluminous ones, and spanning a range well 
outside the peak luminosity--decline rate of the light curve relationship,  
used to make calibrated candles of the SNe Ia. On the theoretical side, 
new explosion scenarios, such as violent mergings of pairs of white dwarfs, 
have been explored. We review those recent developments, emphasizing the new 
observational findings, but also trying to tie them to the different scenarios 
and explosion mechanisms proposed thus far.  
\end{abstract}


\section{Introduction}
\label{}
Type Ia supernovae (SNe Ia) have been the tool that made possible the 
discovery of the acceleration of the expansion of the Universe (Riess et al. 
1998; Perlmutter et al. 1999), and they are now providing new insights on the 
cosmic component, dubbed ``dark energy'', thus revealed. However, in contrast 
with their key role as cosmological probes, and after more than 50 years of 
supernova research, the nature of their progenitors remains elusive. As far 
back as 1960, it was established that Type I supernovae (in fact, the 
now denominated SNe Ia, or thermonuclear supernovae) should result from the 
ignition of degenerate nuclear fuel in stellar material (Hoyle \& Fowler 
1960). The absence of hydrogen in the spectra of the SNe Ia almost 
immediately suggested that they were due to thermonuclear explosions of 
white dwarfs (WDs). Isolated white dwarfs were once thought to be possible 
progenitors (Finzi \& Wolf 1967), but soon discarded due to incompatibility
with basic results from stellar evolution.  
Instead, accretion of matter from a close companion star in a binary system, 
by a previously formed C+O white dwarf with a mass close to the Chandrasekhar 
mass, provides a viable mechanism to induce the explosion (Wheeler \& Hansen 
1971). 

Two main competing production channels are still under discussion nowadays.
One possible path is the so--called single degenerate (SD) channel,
where a C+O white dwarf grows in mass by accretion from a non--degenerate 
stellar companion: a main sequence star, a subgiant, a helium star, a red 
giant, or an AGB star (Whelan \& Iben 1973; Nomoto 1982). Another possible 
path is the double degenerate (DD) channel (Webbink 1984; Iben \& Tutukov 
1984), where two WDs merge due to the loss of angular momentum by 
gravitational radiation. The merging could produce the collapse of the white 
dwarf (Saio \& Nomoto 1985), or it can produce a larger C+O white dwarf 
configuration that then explodes (Pakmor et al. 2012).  

\bigskip

In the decade of the 90's, the variety amongst SNe Ia was discovered, ranging 
from events such as SN 1991bg to those as SN 1991T, through normal SNe Ia
(see Filippenko 1997a,b; Branch et al. 2007; Leibundgut 2011). Such 
diversity was made amenable for cosmology when the correlation of the 
luminosity at the maximum of the light curve of each SN Ia with its rate of 
decline was parameterized (Phillips 1993, 1999; Riess, Press \& Kirshner 1995; 
Perlmutter et al. 1997). It became clear, then, that SNe Ia could 
be used as distance indicators in cosmology, and that led to the 
aforementioned discovery.

Yet, the first decade of the new century has brought new surprises: 
super--Chandrasekhar supernovae, as well as extremly faint ones (see below). 
Neither of them are useful for cosmology, although they are not a severe 
nuisance there, since they can be easily identified, and eliminated from the 
large samples of SNe Ia collected for cosmological probes. Also, various teams 
have started to measure supernova rates at a wide variety of redshifts. The 
idea of using SNe Ia rates to discover the nature of the progenitor systems 
has now become an active line of research. Finally, high--resolution 
spectroscopic observations of SN have yielded the surprising 
result of time--varying absorptions, which indicate the existence of outflows 
in the circumstellar medium surrounding some SN, and points to 
possible nova activity previous to the explosion. An intriguing C II feature
has been identifieed, close to the Si II line typical of SNe Ia, and that has 
led to thinking in two different directions: either the thermonuclear flame 
does not burn the outermost layers of the white dwarf, or maybe C is a 
signature of the merged white dwarf companion of the SN. There are also better 
estimates of the maximum H mass that could be present in the envelopes of the 
pre--SNe, if the explosions were triggered by accretion from a non--degenerate 
companion. There is continued failure to detect H from the radio emission of 
the SNe Ia, and there could be constraints from the X--ray emission as well. 
The task of searching for the companion star in Galactic supernovae has 
already given some definite results, and there are, now, simulations of the 
impact of the SN ejecta on the companion star that can be compared with the 
observations.    

In the following Sections, we present and discuss those new results. In 
Section 2 we briefly review the different models proposed to explain the 
SN Ia phenomenon. Section 3 examines how the Delay Time Distribution (DTD) 
constrains
the possible SN Ia progenitors. In Section 4 we discuss the carbon and 
oxygen absorption features seen, in recent years, in the spectra of SN Ia at
early times, while Section 5 deals with the emission features at late times.
Section 6 discusses the variable blueshifted sodium feature seen in some
SNe Ia. The X--ray constraints are presented in Section 7, and the radio
constraints in Section 8. In Section 9 we report the limits on the 
luminosities of the companions of SNe Ia obtained from pre--explosion 
images. Section 10 deals with the detection of companions throught the early 
light curves of SNe Ia. Section 11 reviews the direct searches for surviving 
companions, in the Galaxy and in the Large Magellanic Cloud. Section 12 deals
with the identification of possible candidates to SNe Ia through  
reconstruction of the orbital evolution of diverse close binary systems
containing white dwarfs. Section 13 addresses the important problem of the
outliers from the peak brightness--decline rate of the light curve 
relationship used to make these SNe calibrated candles for cosmology. Section 
14 deals with the bulk of SNe Ia used for cosmology. We summarize the current 
state of affairs in the last Section.

\section{Models} 

An ideally complete model of a Type Ia supernova should start 
from the formation and subsequent evolution of the binary system assumed 
to originate it, include possible common envelope episodes, mass tranfer 
stages (generally nonconservative), especially those immediately leading to 
the explosion, rotational states of the two stars involved, and finally 
the ignition process and its development into a full thermonuclear explosion
(hydrodynamics and nucleosynthesis). From that, light curves and spectra of 
the emitted light should be computed, for the different stages of the 
outburst (extending until the nebular phase). The characteristics of the 
resulting remnants should be predicted as well. In addition, the frequency of 
the explosions corresponding to the model has to be estimated, for different 
galactic environments. The whole programme involves very diverse domains of 
expertise, so the progress has been disperse. 
  
\subsection{Double-degenerate models} 

Those models were first proposed by Webbink (1984) and by Iben \& Tutukov 
(1984). The most favored scenario, in the latter work, started from binaries 
with component masses in the range 5--9 $M_{\odot}$. They experienced two 
common envelope stages and ended as a pair of C+O white dwarfs, with masses in 
the range 0.7--1 $M_{\odot}$, separated by distances 0.2--0.35 $R_{\odot}$ and 
orbiting each other with periods $P$ between 12 min and 14 hrs. The system 
then losses angular momentum by emission of gravitational waves and the two 
white dwarfs merge on a time scale ranging from 10$^{5}$ to 10$^{10}$ yr 
(merging of binaries due to the emission of gravitational radiation had already
been considered by Tutukov \& Yungelson 1979). The 
merging would occur through disruption of the less massive component of the 
system. That component fills its Roche lobe first (larger radius and 
smaller Roche lobe). The mass transfer, then, should have a runaway character, 
since the more mass the white dwarf losses, the larger its radius becomes (and 
the smaller its Roche lobe). The material of the disrupted white dwarf would 
form a thick disk around the more massive one, which would accrete mass from 
it until reaching the Chandrasekhar limit and explosively ignite C at its 
centre. With the caveat that the effects on the orbit of the two common 
envelope phases could only be roughly approximated, Iben \& Tutukov (1984) 
found that the rate of such mergings might, alone, account for the Galactic 
SN Ia rate.

In the preceding scenario, it was assumed that the only effect of the
accretion of matter by the more massive white dwarf, from the debris of its 
companion, should be growth up to the Chandrasekhar mass. That was soon 
challenged by Saio \& Nomoto (1985), who argued that a very fast mass
transfer ($\dot M \sim 1\times10^{-5}\ M_{\odot}\ yr^{-1}$) would produce an 
off--centre C flash. A C--burning front would then propagate (nonexplosively) 
down to the centre of the white dwarf, changing the chemical composition from 
C+O to O+Ne+Mg along the way. Electron captures on Mg and Ne would subsequently 
make the Chandrasekhar mass smaller than the white dwarf mass, and 
gravitational collapse would ensue. The outcome should thus be the formation 
of a neutron star, rather than a SN Ia explosion. Nomoto \& Iben (1985) further
concluded that the off--centre C ignition would always occur unless the 
mass accretion rate were less than one--fifth of the Eddington limit for an 
isolated white dwarf. Later hydrodynamic simulations (Benz et al. 1990; 
Guerrero et al. 2004) have confirmed that a heavy accretion disk is formed 
around the most massive white dwarf. Whether a SN Ia ensues would depend on 
the mass accretion rate of the white dwarf from the disk, that being 
determined by the viscosity of the latter: a sufficiently low viscosity would 
allow the white dwarf to grow while avoiding the off--centre, non--explosive 
ignition of C (Mochkovich, Guerrero \& Segretain 1997). The issue remains 
open. Another effect of mass acccretion from a massive disk should be the gain 
of angular momentum by the white dwarf. Its consequences have been studied by 
Piersanti et al. (2003a,b) and Tornamb\'e \& Piersanti (2013). They find that 
rotation can stabilize the white dwarf against contraction, even when its mass 
becomes larger than the Chandrasekhar mass. Later, loss of angular momentum by 
emission of gravitational waves allows contraction, until the conditions for 
explosive C ignition are reached at the centre. Recently, 
Lebanon, Soker \& Garcia--Berro (2014) find constraints on the classical DD
type of progenitor from the disk--originated matter around the exploding WD,
as this matter will be shocked  by the SN ejecta and end up in a radiation
implying a larger progenitor radius than observed.  

A different approach is based on the violent merging of a pair of white dwarfs 
(Pakmor et al. 2010). Another possibility is the direct
collision of a pair of white dwarfs. Such collisions could take place 
in environments of high star number density, like globular clusters or 
the Galactic centre (Benz et al. 1989), and also in triple star systems
where a close pair of white dwarfs have their orbits perturbed by the third, 
more distant star (Thompson 2011; Katz \& Dong 2012; Kushnir et al. 2013; 
Dong et al. 2014). In the 
hydrodynamic simulations of Pakmor et al. (2010), the merging of two equal 
mass ($\sim0.9\ M_{\odot}$) white dwarfs produced a subluminous SN Ia. Further 
modeling (Pakmor et al. 2011) has shown that violent mergings where the primary 
mass is $\sim0.9\ M_{\odot}$ can give rise to subluminous SNe Ia for 
unequal masses too, provided that the mass ratio is more than about 0.8. 
Later on, Pakmor et al. (2012) have found that the violent merging of two C+O 
white dwarfs, with masses 0.9 M$_{\odot}$ and 1.1 M$_{\odot}$ (assumed to 
induce the detonation of the C+O mixture) can account for normal SNe Ia 
as well. That has been very recently confirmed by Garc\'{\i}a--Senz et al. 
(2013), who find that the amount of $^{56}$Ni produced in the collisions 
ranges from 0.1 M$_{\odot}$ to 1.1 M$_{\odot}$, thus covering from subluminous 
through normal up to overluminous SNe Ia. They argue, however, that given the  
distribution of white dwarf masses, mostly subluminous events should arise.
On the other hand, Ruiter et al. (2013) find that the brightness distribution
of the explosions produced by violent mergers matches the shape of that 
observed for SNe Ia (although the issue depends on the occurrence of a 
particular phase of mass accretion during binary evolution). Very recently, 
Kromer et al. (2013a) have successfully explained the narrow emission lines 
of [O I] in the late--time spectra of SN 2010lp (a subluminous SN Ia) by the 
violent merger of two C+O white dwarfs, with masses of 0.9 and 0.76 
M$_{\odot}$. Even more recently, Moll et al. (2013) have found that prompt 
detonations following the merging of two white dwarfs can not only reproduce 
both common and overluminous SNe Ia, but also the width--luminosity relation 
on which the use of these supernovae as cosmological distance indicators 
is based. 

A key point, though, is whether there are or not enough close binary white 
dwarf binaries to account for the rate of occurrence of SNe Ia. One approach to
this problem is to look for such systems in the Galaxy. Napiwotzki et al. 
(2007) have reported the results of a systematic radial velocity survey for 
double degenerate binaries as potential progenitors of Type Ia supernovae. 
More than 1000 WDs and pre--white dwarfs  were observed with the VLT. The 
frequency of He WDs is much higher than that of C+O WDs, and they regard
He WD donors as a possible important channel for SNe Ia. Recently, Badenes 
\& Maoz (2012), using multi--epoch spectroscopy of $\sim$ 4000 white dwarfs 
from the Sloan Digital Sky Survey, have determined the white dwarf merger rate 
per unit stellar mass in our Galaxy. They find that the total rate might
well account for the SN Ia rate in the Milky Way and galaxies of the same type, 
but that the rate of merging of pairs of white dwarfs with a total mass 
above the Chandrasekhar mass is only $\sim$1/14 of the total rate. 
So, unless sub--Chandrasekhar mergers can produce SNe Ia, the 
double--degenerate channel should be, at most, a minor contributor to the 
SN Ia phenomenon.

SN Ia models based on the non--violent merging of two white dwarfs, with a 
total mass below
the Chandrasekhar mass, involve a C+O plus a He white dwarf. The He accreted
by the more massive (C+O) white dwarf detonates and the shock wave thus 
generated can either induce the detonation of the C+O layers immediately 
below or converge near the centre and produce a C+O detonation there (Sim et. 
al. 2012). These authors find that the light curves and spectra of such 
explosions do match those observed in SNe Ia. Besides, Ruiter et al. 
(2011) had calculated that if those double--detonation models were able to 
produce explosions similar to SNe Ia, then the sub--Chandrasekhar explosions 
would account for a substantial fraction, at least, of the observed SN Ia rate. 
They also found that the double white dwarf channel involving a C+O plus a He 
white dwarf should have a distribution of delay times (between formation of the 
binary and the SN explosion) spanning from 800 Myr up to the Hubble time. 

\subsection{ Single--degenerate models.}
 
The channel leading to a SN Ia explosion via mass accretion, by a C+O white 
dwarf, from a non--degenerate binary companion, was first modelled by Whelan 
\& Iben (1973), although the idea, as we have seen, already appears in 
Wheeler \& Hansen (1971): a primary with a mass of 1.8--3 $M_{\odot}$, plus a 
secondary with $M \lesssim 0.8\ M_{\odot}$, initially form a system with an 
orbital period between 5 and 9 years. The primary then evolves and becomes a 
C+O white dwarf, with a mass close to the Chandrasekhar mass. The secondary, 
after $\sim$ 10$^{10}$ yr, becomes an AGB star, fills its Roche lobe, and 
transfers mass to the  white dwarf, which then reaches the Chandrasekhar mass 
and explodes. In this first model, the mass of the secondary was chosen to 
explain the occurrence of SNe Ia in elliptical galaxies, long after star 
formation has stopped. Subsequently, a variety of initial binary systems, in 
which mass accretion by the white dwarf can take place at different 
evolutionary stages of its companion (main sequence, subgiant, red giant, 
AGB), and either from Roche lobe overflow or from a stellar wind, have been 
proposed. Also, mass loss by the two components of the binary can take place 
more than once, as well as mass transfer (either conservative or 
non--conservative). The possibility that the companion might have lost its 
hydrogen--rich envelope and become a helium star, at the time of mass tranfer 
to the white dwarf, has also been considered. 

A problem common to both the single--degenerate models and to the 
double--degenerate models, in which the 
conditions for explosive C ignition, close to the centre of the C+O white 
dwarf, are reached when the white dwarf mass has grown to the Chandrasekhar 
mass (or when rotational support has been lost, for masses above such limit), 
is that thermonuclear burning propagation should be subsonic at first 
(deflagration) and, at some point, become supersonic (detonation). If only 
deflagrations were involved, the explosions could just produce a subclass of 
subluminous SNe Ia (Fink et al. 2014). On the other hand, pure detonations of 
Chandrasekhar--mass white dwarfs would burn the C+O mixture to Fe--peak 
elements entirely, in stark contrast with the observations, that show 
intermediate--mass elements at and around maximum light. A  self--consistent 
modeling of the deflagration/detonation transition still remains elusive, 
although steady progress is being made (Woosley 2007; Aspden, Bell \& Woosley 
2010; Schmidt et al. 2010; Woosley, Kerstein \& Aspden 2011).

There is another problem that only concerns single--degenerate models of 
SNe Ia: the possible explosive ignition of the material (hydrogen or helium) 
accumulated on the surface of the accreting C+O white dwarf, before the
latter reaches the Chandrasekhar mass. Also, in the case of hydrogen, 
a high rate of accretion may not lead to explosion but to the formation of 
an extended envelope, which would be ejected by interaction with the 
mass--donor star, that terminating the accretion process.  

Accumulation of hydrogen on the surface of a white dwarf at a low rate 
should produce nova--like explosions, in which all the accreted material 
(and maybe even more) would be expelled (Nomoto 1982; Livio \& Truran 1992).
The lower limit on the mass--accretion rate, to avoid explosions and strong
flashes, is uncertain and depends on the mass of the white dwarf, but a 
value $\dot M \sim 5\times10^{-8}\ M_{\odot} yr^{-1}$ is often given, although
Kercek, Hillebrandt \& Truran (1999), in 3D simulations, obtained steady 
hydrogen burning for rates as low as $\dot M \sim 5\times10^{-9}\ M_{\odot} 
yr^{-1}$ (see discussion in Hillebrandt \& Niemeyer 2000).  

Even if hydrogen is burned steadily or in weak flashes, the resulting helium 
layer should explosively burn, in a detonation, if the rate of accumulation
of helium were $\dot M \lesssim 5\times10^{-8}\ M_{\odot} yr^{-1}$ (although 
this is only true for white dwarf masses $M \lesssim 1.13\ M_{\odot}$). The 
same applies to the direct accretion of helium. Lower rates would thus be 
excluded from the production of Chandrasekhar--mass explosions, but they 
could lead to sub--Chandrasekhar, edge--lit explosions, instead. 

The explosion of helium shells in accreting white dwarfs as a Type Ia  SN 
mechanism was first proposed by Taam (1980) and Nomoto (1982a)
(the ``double detonation'' model), and numerically studied, in 
one--dimensional (1D) numerical simulations, by Woosley, Taam \& Weaver 
(1986), and by Livne (1990). The earliest 2D simulations were made  
by Livne \& Glasner (1991). Such explosions were further 
investigated by Woosley \& Weaver (1994), and by Livne \& Arnett (1995).
The ``robustness'' of the double detonation model has been recently 
checked in 2D and 3D simulations by Moll \& Woosley (2013), whilst the 
conditions for producing helium detonations have been examined by Woosley 
\& Kasen (2011). The first evidence of a sub--Chandrasekhar explosion was 
found by Ruiz--Lapuente et al. (1993) (later confirmed by Mazzali et 
al. 1997), from modeling the spectra of SN 
1991bg. Whether double detonation models  produce explosions characteristically 
similar to those of SNe Ia remains an open question, but Ruiter et al. 
(2011) find that the helium star channel would have delay times $<$ 500 
Myr (``prompt explosions''), while the double white dwarf channel (C+O 
plus He white dwarf) would have longer delay times, as mentioned above.

Successful accretion of hydrogen, leading to growth of a C+O white dwarf up
to the Chandrasekhar mass through steady shell hydrogen and helium burning, 
thus requires high accretion rates. That can be achieved for binary stellar 
companions at different stages ot their evolution, but if hydrogen 
accumulates faster than it can be burned, a red giant--like structure should
form, soon engulfing the companion and leading to a common--envelope stage. 
A solution to this problem was found by Hachisu, Kato \& Nomoto (1996): 
when the mass accretion rate exceeds the maximum rate at which hydrogen can be 
burned, there is no static envelope solution, and the excess material is 
blown off in a wind. The optically thick wind solution had been previously
found to explain the light curves of nova outbursts by Kato \& Hachisu (1994). 

White dwarfs accreting mass at high rates should emit large amounts of 
radiation in the X--ray band, and they should appear as luminous supersoft
X--ray sources (van den Heuvel et al. 1992; Di Stefano \& Nelson 1996;  
Yungelson et al. 1996; Kahabka \& van den Heuvel 1997; Li \& van den 
Heuvel 1997;  Orio 2006). However, as we will see in Section 7, 
that might be in conflict
with the observed X--ray emission of elliptical galaxies and galaxy bulges.

Another possible channel involving a C+O white dwarf plus a non--degenerate 
companion star is the core--degenerate scenario recently advocated by Soker 
(2013): a Chandrasekhar or super--Chandrasekhar white dwarf is formed 
from the merging of a white dwarf with the hot, more massive core of an AGB 
star. The initial white dwarf is disrupted and its material accreted by 
the AGB core, that leading to the formation of a rapidly spinning, more
massive white dwarf. The delay time till explosion would then be given by 
the spin--down time of the new white dwarf. Mergings of a white dwarf with the
core of an AGB star, following a common--envelope episode, had also been 
considered as a production mechanism of SNe Ia by Sparks \& Stecher (1974) 
and by Livio \& Riess (2003). Soker et al. (2013, 2014) have proposed
the core--degenerate scenario to explain the characteristics of two very 
different SNe: PTF11kx and SN 2011fe. 

In the preceding, we have not dealt with the especifics of the hydrodynamic 
modeling of the different types of explosions surveyed, which has reached 
unprecedented standards of realism (see Hillebrandt et al. 2013). They are
being matched by 3D models of the spectra that should arise from the 
explosions (Baron, Hauschildt \& Chen 2009).

\section{Constraints on progenitors from DTD}

The Delay Time Distribution (DTD) is given by the time
evolution of the SN rate that would follow an instantaneous
burst of star formation. It is related to the observed rate
$r_{SNe Ia}$ by:
\begin{equation}
r_{SNe Ia} (t) = \int_{0}^{t} R(t - \tau)SFR (\tau) d\tau
\end{equation}
where $R(t)$ is the DTD, $SFR$ is the star formation rate, 
and $t$ and $\tau$ are in the SN rest frame (see, for instance, 
Ruiz--Lapuente \& Canal 1998).

Early research on SNe Ia rates and DTDs indicated the need of a 
two--component model at least: one that could be fitted with a short DTD 
population ($\sim10^{8}$ yr) and another one with a long DTD 
population (3--4 Gyr) (Scannapieco \& Bildsten 2005; Mannucci et al. 
2006; Brandt et al. 2010). Recent studies (Maoz \& Mannucci 2012) 
suggest that the DTD peaks at the  shortest times, as a function  
$\sim t^{-1}$ (see also Mannucci et al. 2005, and Oemler \& Tinsley 1979). 
Such a function characterizes the dominance of double 
degenerate mergings. As already mentioned in the previous Section, Badenes 
\& Maoz (2012) find that the total merger rate of white dwarf pairs 
(for sub--Chandrasekhar WDs) 
is similar to the observed SNe Ia rate (see also Ruiter et al. 2011).

The SNe Ia searches in clusters of galaxies at high redshifts (Barbary et al. 
2012) have indicated a $t^{\alpha}$ behavior of the rates, with $\alpha$ $=$ 
-1.41$^{+0.47}_{-0.40}$, which also favors the WD merging channel.
Such best fit value is consistent with measurements of the 
late DTD in field galaxies (Totani et al. 2008). Most predictions 
for the SD scenario show a steeper late--time DTD
(Greggio 2005; Ruiter et al. 2009; Mennekens et al. 2010), 
whith  $\alpha$ ranging from $\alpha$ $=$ -1.6
(Greggio 2005) to $\alpha$ $<$ -3 (Mennekens et al. 2010), 
depending on the details of the scenario and on how the binary evolution 
is calculated (see,
however, Hachisu et al. 2008, and Pritchet et al. 2008). Recently, 
Bonaparte et al. (2013) have computed the cosmic SN Ia rate, for several 
cosmic star formation rates and progenitor models, and compared it with 
the observational data. No firm conclusions can be derived, concerning the 
SN Ia progenitors, but the existence of prompt SNe Ia, exploding within the 
first $10^{8}$ yr after the corresponding systems are formed, is required, 
althought their fraction should not exceed 15--20\% of the total, to be 
consistent with the chemical evolution of the galaxies (see, for an updated
view of this subject, the review by Maoz, Mannucci \& Nelemans 2014). 
Also, from the analysis of the environments of 90 Hubble flow SNe Ia discovered
by the {\it Nearby Supernova Factory}, Rigault et al. (2013) find evidence
of two distinct populations with different ages: one associated with current
star formation and another one corresponding to passively evolving 
environments. 

\section{Carbon and oxygen absorption features at early times}

If the SNe Ia explosions would come from white dwarf mergings,
one would expect to see carbon and oxygen in the early--time spectra, 
coming from the surrounding clumps of this material, originated by the merging.
With the access to large numbers of SNe Ia, the {\it SN Factory} has
found C II absorption lines in the spectra of several of them (Thomas
et al. 2011). These authors estimate that 22$^{+10}_{-6}$ $\%$ of SNe Ia
exhibit  C II signatures as late as 5 days before maximum light.
In some  cases one can treat them as spherically symmetric 
absorptions, in others as ``carbon blobs''. 
In the context of explosions from the merging of two C+O white dwarfs, 
the presence of this photospheric carbon at high velocities seems 
justified.  Parrent et al. (2011) have studied both the 
spherically--symmetric C II absorption and the non--spherically 
symmetric cases. Altavilla et al. (2007) found evidence of a C blob
in the normal SN Ia SN 2004dt. Folatelli et al (2012) have also
found evidence of unburnt carbon in SNe Ia from the {\it Carnegie Supernova 
Project}, in 30$\%$ of the objects (see Figure 1).

In the case of the normal SN 2011fe, in the M101 galaxy, the availability of
early spectra has allowed to see both C II and O I emissions (Parrent 
et al. 2012; Nugent et al. 2011). The 
absorption of O I appears at higher velocities, which suggests that SN 2011fe 
may have had an appreciable amount of unburned oxygen within the 
outer layers of the ejecta. Mazzali et al. (2014), from modeling of the
spectral evolution, find that the high--velocity tail of the ejecta  
differs from the predictions of both deflagration and delayed--detonation
models of the explosion. 

\begin{figure}
\centering
\includegraphics[width=0.6\columnwidth]{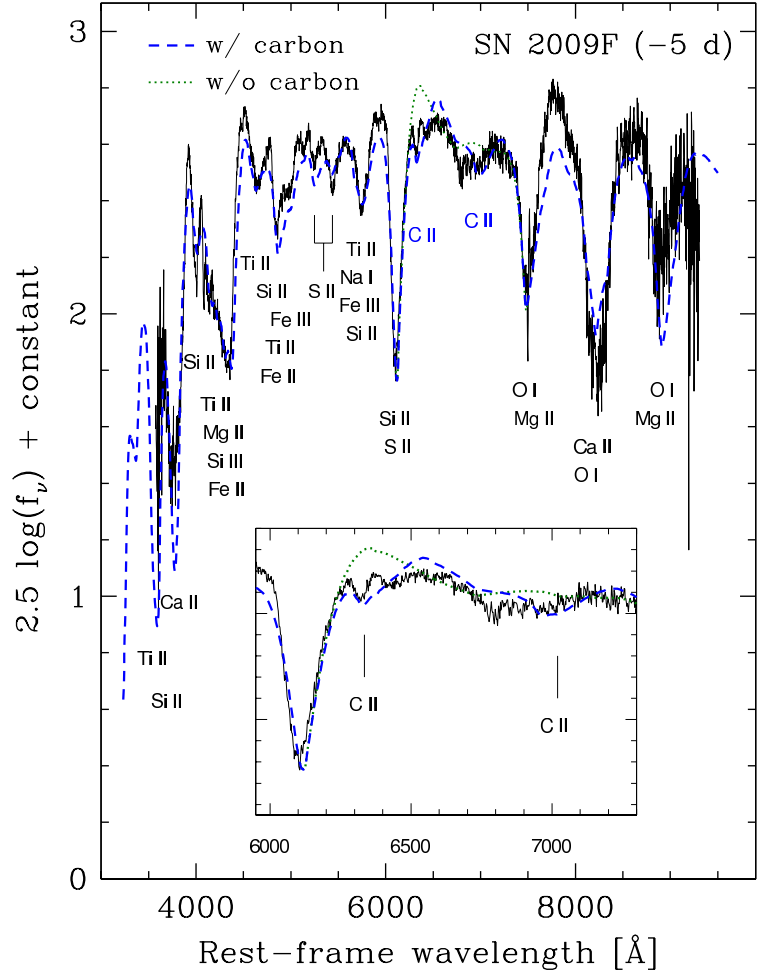}
\caption{Spectrum of SN 2009F at -5 days (black solid line) showing 
the C II $\lambda$6580  and $\lambda$7234 lines, and a matching synthetic
spectrum from a model with carbon (dashed line). A synthetic spectrum 
from a model without carbon  is shown in dotted lines. (See Folatelli 
et al. 2012). (Courtesy of Gaston Folatelli. \copyright AAS. Reproduced with 
permission).}
\label{Figure 1}
\end{figure}

Nomoto, Kamiya \& Nakasato (2013) suggest that unburnt C inside 
the ejecta might just come from asymmetric SNe Ia explosions, in a  
single--degenerate model, for instance, if off--center ignitions take place 
(Maeda et al. 2010b). However, 
it is also possible that unburnt C would be left inside the ejecta
in explosions coming from mergings of C+O WDs, even from violent ones. 
Such carbon would not have been ignited by the detonation that takes place
in such mergings (Hicken et al. 2007; Pakmor et al. 2010; Hillebrandt et al.
2013). In that case, we would have a minimum of $\sim$30 $\%$ of SNe Ia 
coming from double--degenerate systems. 

It is important to study through spectropolarimetry (Wang and Wheeler 2008;
H\"oflich 1991 ; Jeffery 1989, 1990 ) the asymmetry of the C II and O I 
features. Thus far, line asymmetries have been seen in some SNe Ia, such as 
SN 2004dt (Wang and Wheeler 2008).

\section{Emission features at late times}

\begin{figure}
\centering
\includegraphics[width=0.6\columnwidth]{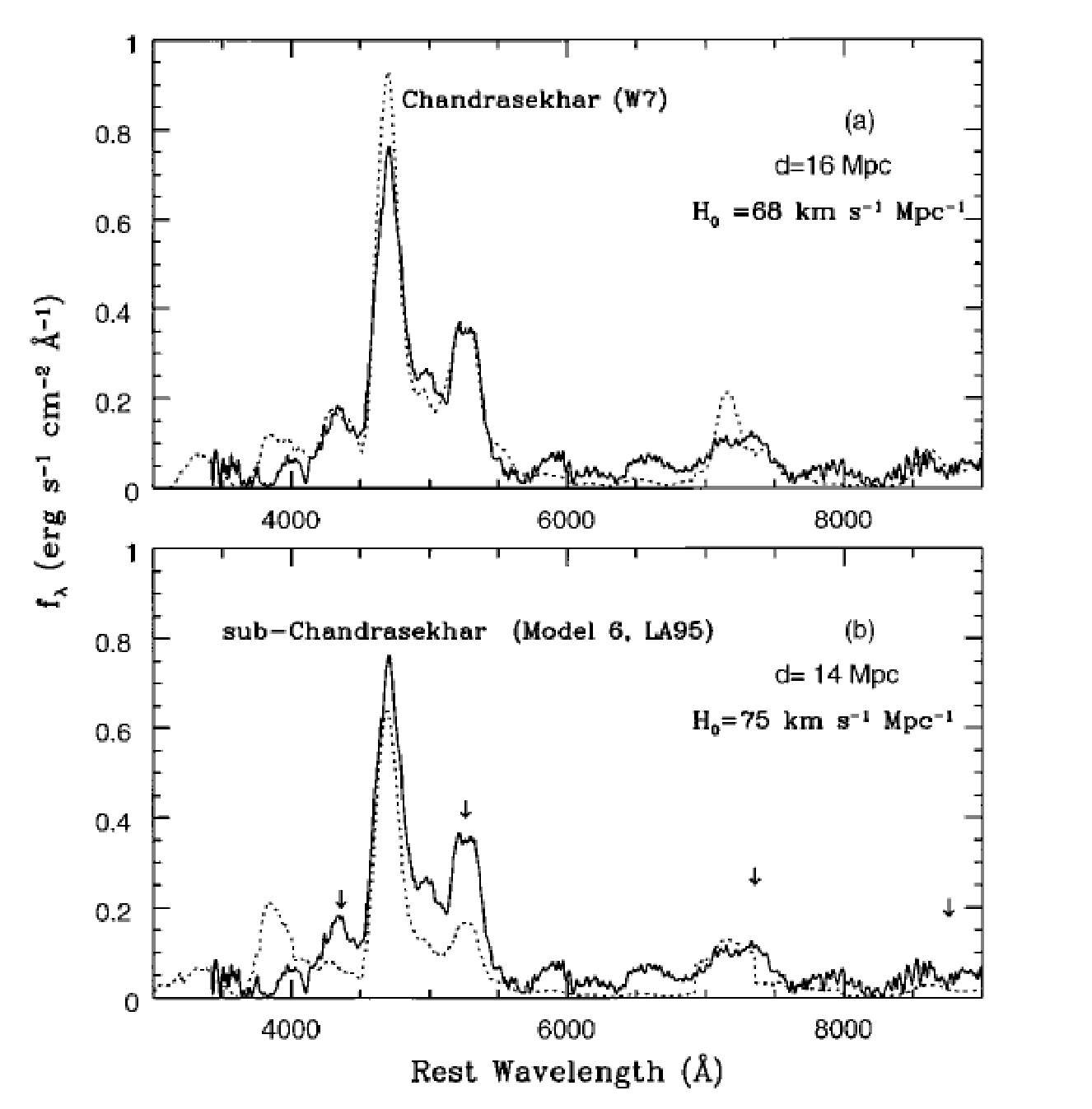}
\caption{Chandraskhar models were favored by normal nebular SNe Ia. 
Sub-Chandrasekhar He--detonations (Livne \& Arnett 1995) were 
largely disfavored (Ruiz--Lapuente 1996. \copyright AAS. Reproduced with 
permission).}
\label{Figure 2}
\end{figure}

The merging models of C+O WDs need testing in the nebular phase.
As discussed in R\"opke et al (2012), several properties of the
merging models could be observable in that phase. In general, the
model of (initial) deflagration of a WD when reaching the Chandrasekhar mass 
leads to a high degree of neutronization, since burning occurs at
densities higher than $2\times10^{8} g\ cm^{-3}$.
Such neutronization can be seen, for instance, from the nebular emission
of Ni II (coming from the stable isotope $^{58}$Ni). In the violent
merging of two white dwarfs, instead, the burning occurs at peak densities
below $2\times10^{8} g\ cm^{-3}$ (R\"opke et al. 2012), and
nebular emission of stable Ni should not be so prominent. In fact,
the nebular emission of the decay products of $^{56}$Ni
is expected to be asymmetric, both in the single degenerate model and in the 
non--violent merging of two C+O WDs, the asymmetry arising from the material 
produced by a deflagration initiated off--centre, when the WD reaches the 
Chandrasekhar mass. The Ni produced later, in the detonation phase, would 
be more symmetrically distributed, instead. Those asymmetries of the nebular 
emission by the products of $^{56}$Ni decay have already been seen, according 
to Maeda et al. (2010a,b). Maeda et al. (2010a) assert that the nebular 
spectra do reveal ignitions offset from the centre of the WD, and 
that this is a generic feature of SNe Ia. 

A point made by R\"opke et al. (2012) is that, in the violent  
merging of C+O WDs, the detonation of the secondary WD, at low densities, 
should introduce copious amounts of oxygen in the innermost ejecta. That could 
give rise to visible [O I] $\lambda\lambda$ 6300, 6364 \AA.
This has been seen in  SN 2010lp (Kromer et al. 2013a; Taubenberger et al. 
2013a), and it gives strong support to that scenario.
A different question (Ruiz-Lapuente 1996) is that, at lower densities, the 
[Fe II] and [Fe III] lines are less collisionally excited and become weaker 
than what is seen in the data. This was shown for the sub--Chandrasekhar
 edge--lit detonations of Livne \& Arnett (1995) which gave a very poor fit
to the observations, while Chandraskehar models with 0.6 M$\odot$ of $^{56}$Ni 
fit very well the data
of normal SNe Ia. The fit to the nebular data for the 
new class of sub--Chandrasekhar explosion models remains to be tested, since 
the result just mentioned came from the models of He detonations in 
sub--Chandrasekhar WDs available in the 90's 
(Ruiz--Lapuente 1996, hereafter R96; see Figure 2).

On the other hand, in the models based on Chandrasekhar--mass WDs resulting 
from accretion of H from a non--degenerate star, one would 
expect to see emission of H at $\lambda$6563 \AA. 
Such emission is not seen in normal SNe Ia (Leonard 2007). 
Hydrodynamic simulations indicate that seeing the H$\alpha$ emission is 
related to the mixing of H with other elements. In a hydrodynamic simulation 
for a main--sequence companion to the SN (Liu et al. 2012), up to 19$\%$ of 
the total mass of the companion star is stripped by the impact of the SN 
ejecta, and those debris should mix with the most slowly moving layers of the 
ejecta. 

There have been a few SNe Ia with H$\alpha$ in emission.  
Such are, for instance, the cases of SN 2002ic (Hamuy et al. 2003),
of SN 2005gj (Aldering et al. 2006), and of PTF11kx (Dilday et al. 2012). 
We will address this point in Section 13.2.

\section{Variable blueshifted Na I D features}

As first seen by Patat et al. (2007), some SNe Ia have variable Na I D 
absorption lines, significantly  blueshifted with respect to the absorption 
features of other elements. This result has been confirmed by Simon et al. 
(2009), Sternberg et al. (2011), and Foley et al. (2012) (see Figure 3). The 
general 
interpretation is that a significant fraction of SNe Ia progenitor systems 
have outflows of material previous to the explosion. The Na I D lines arise 
from the ionized circumstellar medium (CSM). 
Foley et al. (2012) find a correlation with higher velocity ejecta in the 
SNe Ia that show blueshifted Na I D line profiles. They suggest the possibility
that progenitor systems with strong outflows tend to have more kinetic 
energy per unit mass than those with weak or no outflows. 

Patat et al. (2013) have shown that the SN 2011fe was surrounded by a 
``clean'' environment, and there is a lack of time--variable blueshifted 
absorption
features. They found SN 2011fe consistent with the progenitor being a binary 
system with a main--sequence or even a degenerate star. 
Nugent et al. (2011) found that the exploding star was likely a C+O WD and, 
from the lack of an early shock, that the companion was most likely a 
main--sequence star or there is no surviving companion. Li et al. (2011), 
from pre--explosion images, also 
exclude companions more evolved that subgiants (see Section 9). 
Bloom et al. (2012) also find that only degeneracy--supported 
compact objects---WDs and neutron stars---are viable as the primary star. With 
few caveats, they also restrict the companion (secondary) star 
radius to $R_{c} \leq 0.1 R_{\odot}$, that excluding Roche--lobe 
overflowing red giant and main--sequence companions to high significance.

It would be interesting to see if, within the sample of SNe Ia 
showing C II and O I absorptions, there are cases of outflows. It is tempting 
to think that the supernovae with variable Na I D features are connected to 
nova precursors and the ones showing C and O in the outermost layers are 
connected with mergings of WDs, instead. 

Recently, Sternberg et al. (2013) have found that 18\% of the SNe Ia events
show time--variable Na I D features  associated with circumstellar material.
One might tentatively associate them with recurrent novae.

\begin{figure}
\centering
\includegraphics[width=0.6\columnwidth,angle=-90]{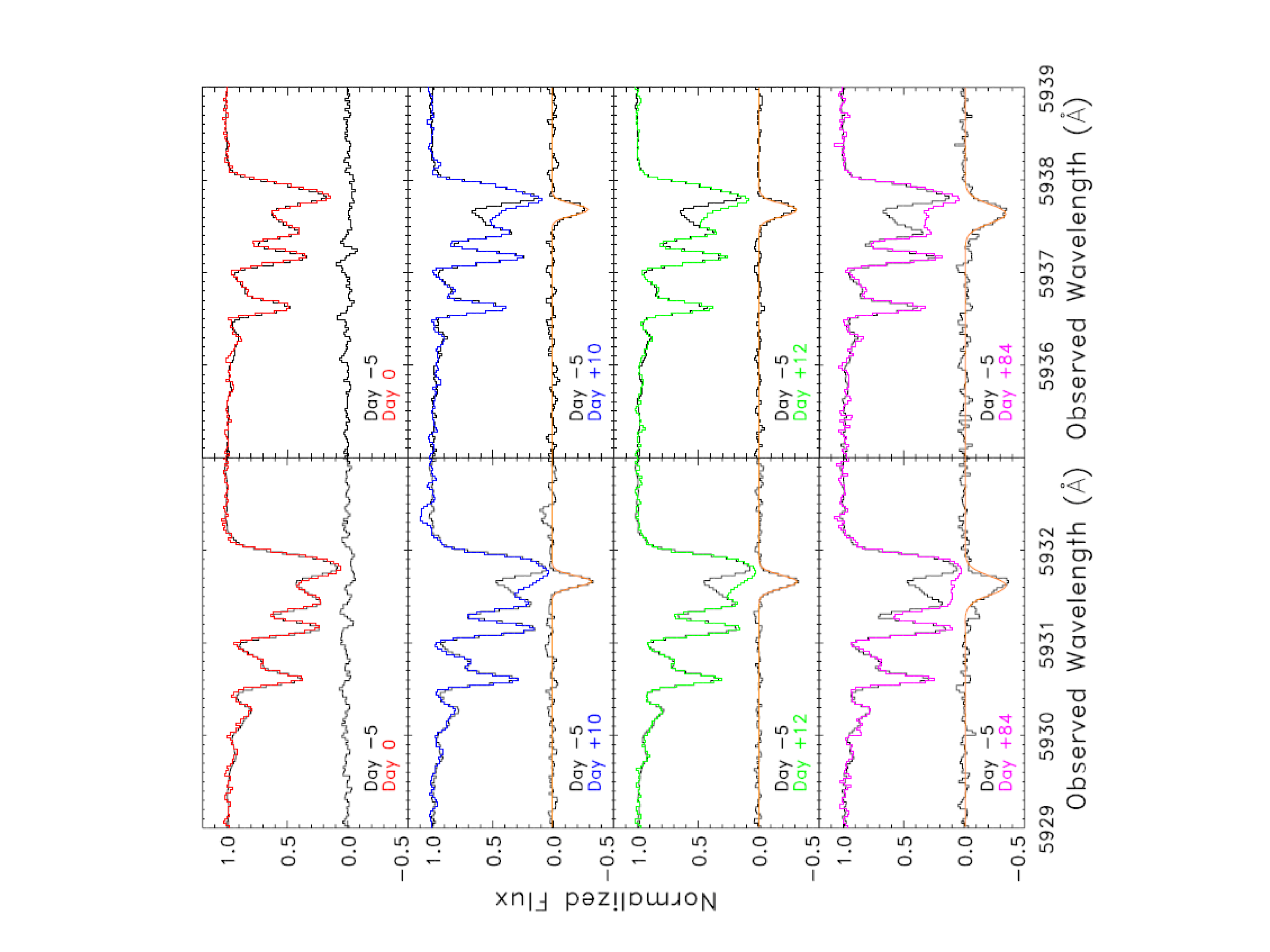}
\caption{High--resolution observations of the Na D absorption lines in the
spectrum of SN 2007le (day -5 and subsequent days). Each panel compares 
with the previous observation  and also with the first one (dotted lines). 
One can see that the Na absorption line, at $\lambda$ 5931.5 \AA, strengthens 
with time. (Courtesy of Josh Simon; Simon et al. 2009. \copyright AAS. 
Reproduced with permission).}
\label{Figure 3}
\end{figure}
 
\subsubsection{Recurrent novae as progenitors of SNe Ia}

Recurrent novae are binaries harboring a WD close to
the Chandrasekhar mass. Classical novae are the outcome 
of unstable thermonuclear burning in accreting white dwarfs.
Those white dwarfs have typical masses $M \simeq 0.8\ M_{\odot}$.
If the accretion rate and white dwarf mass are high, the 
time between flashes can become short enough that the recurrence 
can be observed. Due to the large accretion rate and insignificant 
mass loss by ejection, it has been proposed that in recurrent novae (RNe)  
the white dwarfs may grow to the Chandrasekhar mass and give rise to 
SNe Ia (Starrfield et al. 1988;  Schaefer 2010).

Indeed, RNe can be the progenitors of a part of the SNe Ia
that show variable circumstellar Na absorptions. 
Amongst recurrent novae, there are those of the U Sco type,  
where the donor is a main sequence or subgiant star, and those of the
RS Oph type, where the donor is a red giant. 
The supernova PTF11kx,  classified as a SN Ia--CSM for its interaction 
with the circumstellar medium, is believed to have a symbiotic recurrent 
nova as its progenitor system (Dilday et al. 2012), although it has also been 
attributed to a violent prompt merger of a white dwarf with the core of 
a massive AGB star by Soker et al. (2013), who also argue that the mass of
the shell surrounding PTF11kx is too high to have been produced by a recurrent
nova. They estimate the hydrogen mass in the shell to be 
$M_{sh} > 0.2 M_{\odot}$. Also, L\"u, Yungelson \& Han (2006) have shown that
symbiotic novae are unlikely SNe Ia progenitors, due to their low efficiency
in hydrogen accumulation.

Sahman et al. (2013) suggest that the recurrent nova Cl Aql will become a 
SN Ia within 10 Myr. They find that the mass 
of the white dwarf is 1.00 $\pm$ 0.14 M$_{\odot}$, and the mass 
of the companion is 2.32 $\pm$ 0.19  M$_{\odot}$. The radius of
the latter is 2.07$\pm$ 0.06 R$_{\odot}$. They estimate that 
the secondary is a slightly evolved A--type star, and suggest that the 
system is rapidly evolving into a supersoft X--ray source. 
Patat (2011) sees, in the variable sodium lines of some SNe Ia, 
a possible connection with recurrent novae.

Recently, Soraisam \& Gilfanov (2014) have compared the nova statistics for
M31 with the SNe Ia rates. They find that significant mass accumulation, in
the unstable burning regime, is only possible for WDs with masses below 1.25 
$M_{\odot}$. More massive WDs do not significantly accumulate mass. Thus, the
final stage of mass growth can not occur at low mass--accretion rates, when 
the burning is unstable. Therefore, to be SNe Ia progenitor candidates, the 
systems should go into the stable burning regime in the final phases.

\subsubsection{Other wind--blowing systems made of white dwarfs with 
main--sequence or subgiant companions}

Systems consisting of a mass--accreting white dwarf and a Roche--lobe filling, 
more massive, slightly evolved main--sequence or subgiant star, steadily 
burning H, should appear as luminous supersoft X--ray sources (see Section 2). 
With varying mass--accretion rates, however, they can also burn H unstably,
at times, and then appear as recurrent novae of the U Sco type. Nomoto et al. 
(2002) (see also Hachisu et al. 1999) propose an scenario in which a C+O white
dwarf is formed from a red--giant star with a helium core of $\sim 0.8-2.0\ 
M_{\odot}$. Following a first common--envelope episode, a helium star results 
and then evolves to form a C+O white dwarf of $\sim 0.8-1.1\ M_{\odot}$. A 
part of the helium envelope would have been transferred to the main--sequence
companion. The white dwarf would thus accrete and burn a mixture of H and 
helium. Depending on the mass--accretion rate, a wind might be blown from the 
surface of the white dwarf. It should not be optically thick enough to absorb 
all the X--ray emission, but it could, however, absorb a part of the soft X rays
(Hachisu, Kato \& Nomoto 2010; see next Section). It will be intersting to see
whether variable circumstellar material would be observed in this scenario.

\section{ X--ray constraints}

The two different channels to SN Ia explosions, the single--degenerate path 
and the double--degenerate one, lead to 
very different predictions for the X--ray emission (Gilfanov \& Bogd\'an 
2010). Whereas no strong X--ray emission is expected, prior to explosion,  
in the merger scenario, in the single--degenerate scenario the 
white dwarf that accretes mass from a non--degenerate companion becomes a
source of X--rays for about 10$^{7}$ yr before the explosion.
If the growth in mass of the white dwarf is due to accretion of hydrogen,
followed by steady burning of hydrogen into helium, one expects a 
thermonuclear luminosity
\begin{equation}
L_{nuc} \sim \epsilon_{H}X\dot M\ {\rm erg\ s^{-1}}
\end{equation} 
where $\epsilon_{H}$ is the energy per unit mass released by hydrogen 
burning, $X$ is the hydrogen mass fraction in the accreted material, and 
$\dot M$ is the mass--accretion rate. For standard values. 
$L_{nuc} \sim 10^{37}$ erg s$^{-1}$, which is more than one order of magnitude
larger than the gravitational energy released by accretion, 
$L_{grav} = GM\dot M/R$ ($M$ and $R$ being the mass and radius of the
white dwarf). That sustains a surface temperature of the white dwarf:
\begin{equation}
T_{\rm eff} \simeq 
67\left({\dot M\over 5\times10^{-7}M_{\odot}/{\rm yr}}\right)^{1/4} 
\left({R_{\rm WD}\over 10^{-2}R_{\odot}}\right)^{-1/2}\ {\rm eV}
\end{equation} 
Such sources are observed in the Milky Way and nearby galaxies, and they 
are known (see Section 2.2) as supersoft sources 
(van den Heuvel et al. 1992; Kahabka \& van den Heuvel 1997).
Gilfanov \& Bogd\'an (2010) report that the observed X--ray flux from
six nearby elliptical galaxies and galaxy bulges is a factor 
$\sim$30--50 less than predicted by the accretion scenario, based upon an
estimate of the supernova rate. They conclude that no more than $\sim$
5\% of Type Ia supernovae in early--type galaxies can be produced by white 
dwarfs 
accreting hydrogen in binary systems. Hachisu, Kato \& Nomoto (2010), 
however, suggest that there is, in fact, no inconsistency, since symbiotic 
supersoft sources have fluxes $\sim 0.4\times10^{36}$ erg s$^{-1}$ in the  
0.3--0.7 kev range. There is also uncertainty in theoretically deriving the  
X--ray luminosity of the supersoft sources, due to the still rough atmosphere 
models of mass--accreting WDs and to the neglect of absorption of the soft 
X--rays by the cool wind material from the companion star.  

On the other hand, X--ray emission can inform us about the circumstellar
medium around the SNe Ia (Badenes et al. 2007). In that work, the authors
disfavor optically thick accretion winds from the WD surface. Such 
winds would produce large cavities in the interstellar medium (ISM).
The fundamental properties of the seven  supernova remnants (SNRs) of type Ia
of their sample (SN 1885, Kepler, Tycho, SN 1006, 0509-67.5, 0519-69.0, and  
N103B) are incompatible with SNR models expanding inside such cavities.
In general, the search for X--ray emission at the time of the supernova
outburst has also provided probes of the circumstellar medium, which so far is 
considered to be of low density (Hughes et al. 2007).  Recently,
Rusell \& Immler (2012) have examined 53 SNe Ia observed 
with the {\it Swift} X--ray telescope, and their upper limit
to the X-ray emission gives further evidence that the companion
stars in SNe Ia are neither massive nor evolved (post main--sequence), 
due to the corresponding limit on wind mass--loss rate inferred. They can 
not rule out, instead, 
main--sequence star companions, with mass--loss rates $< 10^{-7} M_{\odot} 
yr^{-1}$. A double white dwarf system is also 
permitted, due to the lack of circumstellar interaction and hence lack of 
X--rays there. 

The tightest constraint on the progenitor of a SN Ia, coming from 
X--ray emission, is that for SN 2011fe (Horesh et al. 2012; Margutti et al. 
2012). The X--ray observations yield an upper limit 2$\times$10$^{-9}$ 
M$_{\odot}$ yr$^{-1}$ to the mass outflow (assuming a wind velocity 
$v_{wind}$ = 100 km s$^{-1}$). As we have seen in 
Section 2, accretion at a rate $\sim$ 10$^{-7}$ M$_{\odot}$ yr$^{-1}$ is 
thought to be necessary for stable accretion and nuclear burning on the 
surface of a white dwarf (Nomoto 1982). Supersoft sources can achieve those 
rates (Kahabka \& van den Heuvel 1997), although the rates can also be 
either lower or higher, in these sources. Horesh et al. (2012) analyse the 
models of interaction of the wind with the circumstellar material and conclude 
that the data from SN 2011fe can rule out a symbiotic system, but not 
a main sequence or subgiant mass--donor. The same conclusion is reached from 
their analysis of the radio emission (see next Section). Margutti et al. (2012)
also discard symbiotic systems, as well as Roche lobe overflowing subgiants and 
main--sequence secondary stars if $\geq$ 1\% of the transferred mass is lost at 
the Lagrangian points.  

The nearby SN 2014J  has provided a new opportunity to test the presence of
X-ray emission from the pre--explosion X--ray images (Nielsen et al. 2014).
According to these authors, the upper limits from the  {\it Chandra} 
X--ray observatory do exclude a classical super--soft source as the progenitor. 
Near the Chandrasekhar mass, the effective temperature corresponding to the 
stable nuclear burning on the WD surface exceeds 100 eV. For this temperature, 
the 3\,$\sigma$ upper limit on the bolometric luminosity is 
$\simeq 3.8 \times 10^{37}\ erg\ s^{-1}$, assuming a column density of 
hydrogen $N_{H} = 6.9 \times 10^{21}\ cm^{-2}$ and a black body spectrum. That 
confidently excludes a classical super--soft source during the final stages 
of the mass accumulation by the progenitor. Due to the large absorption, the 
{\it Chandra} upper limits are less constraining at lower temperatures. They 
do not exclude, therefore, less conventional progenitors, e.g. a WD enshrouded 
in an optically thick envelope or wind. Deep X--ray observations of the 
post--explosion environment (Margutti et al. 2014) now rule out 
single--degenerate progenitors with steady mass loss until the time of the 
explosion (the maximum mass spilled by the system should be $\leq$ 1\%), and 
do only allow recurrent novae with a recurrence time $<$ 300 yrs, stars where 
the mass loss ceases before the explosion, or double WD systems. 

From a different angle, there have been, from the start, great expectations 
to detect hard X--ray and $\gamma$--ray photons from SN 2014J (Isern et al. 
2013; The \& Burrows 2014), since the supernova was close enough to be detected 
by {\it INTEGRAL} and  {\it NuSTAR}. It has, indeed, been detected by 
{\it INTEGRAL} (Churazov et al. 2014). The line flux suggests that 0.62 $\pm$ 
0.13 $M_{\odot}$  of radioactive $^{56}$Ni have been synthesized in the core.
 The mass of 
the ejecta (from the continuum emission) would be $\sim$ 1.4 $M_{\odot}$ 
and composed of roughly equal fractions of iron--group and intermediate--mass 
elements. There is thus agreement with the model of the explosion of a 
Chandrasekhar--mass WD. Diehl et al. (2014) find that about 0.06 
M$_{\odot}$ of $^{56}$Ni should be at the outskirts of the ejecta. This has
suggested that He accreted by the white dwarf could have exploded in the
external layers and triggered the central ignition. 

\section{Radio emission}

The lack of radio emission from Type Ia supernovae has been useful 
in discarding one type of single--degenerate path as a major contributor: 
SNe Ia from 
symbiotic systems. In symbiotic systems, the white dwarf accretes mass from 
the wind of a giant or AGB companion. The wind accretion should produce 
radio emission when the SN ejecta interact with the circumstellar environment 
created by such systems. Panagia et al. (2006) set upper limits on mass--loss
rates of $\sim$ 10$^{-7}$ M$_{\odot}$ yr$^{-1}$. Hancock et al (2011)
suggest upper limits, to the average mass--loss rate of the companion by 
stellar wind, of $1.3\times10^{-7} M_{\odot} yr^{-1}$. These authors say that
such limit is inconsistent with SNe Ia in which the accretion comes 
from intermediate or high--mass companions. Instead, a main sequence star 
having fast winds ($\>>$ 10 km $^{-1}$) could remain undetected, even with 
much higher mass--loss rates. 

The nearby supernova SN 2011fe has made possible the most sensitive radio 
study of a SN Ia made up to now (Chomiuk et al. 2012). The data set direct 
constraints to the density of the surrounding medium at radii 
$\sim 10^{15}-10^{16}$ cm, that implying an upper limit on the mass--loss rate 
from the progenitor system of 
$\dot{M} \lesssim 6\times 10^{-10}\ {{M}_{\odot }\ yr^{-1}}$ (assuming a 
wind speed of 100 km s$^{-1}$), or expansion inside a uniform CSM with 
density $\lesssim$ 6 cm$^{-3}$. Drawing from the observed 
properties of non--conservative mass transfer in accreting white dwarfs, 
they use the limits on the density of the circumstellar environment to 
exclude a good fraction of the parameter space of possible progenitor systems 
of SN 2011fe. A symbiotic progenitor system can be ruled out, as well as any 
other system characterized by a high mass--transfer rate onto the white dwarf  
which could give rise to optically thick accretion winds. Assuming that a 
small fraction, $\sim$ 1\% of the mass transferred, is lost from the 
progenitor system, they can also eliminate much of the parameter space 
occupied by potential progenitors such as recurrent novae or, alternatively, 
progenitors undergoing 
stable nuclear burning. They eliminate, therefore, for SN 2011fe, a large 
fraction of the parameter space associated with popular single--degenerate 
progenitor models, leaving only a limited region, mostly inhabited by some 
double degenerate systems, as well as by exotic single degenerates in which a 
sufficient time delay takes place between mass accretion and SN explosion. 

The even closer SN 2014J has also been observed in radio with the VLA, 
without any detection (Chandler \& Marvil 2014).  This points out to a 
surrounding medium of low density as well.

\section{Limits from pre--explosion images}

It has been possible to put constraints on the
progenitors from  pre--explosion images in other galaxies. 
This endeavour has been of particular interest for SN 2011fe, since 
it exploded in the galaxy M101, at 6.4 Mpc only. Another nine SNe Ia 
with preexisting {\it HST} data on their host galaxies have also been close
enough (within 25 Mpc) to search for the progenitors. It has only been possible 
to set upper limits which rule out normal stars with initial masses
larger than 6 M$_{\odot}$ at the tip of the AGB branch, 
young post--AGB stars with initial masses larger than 4 M$_{\odot}$, and 
post--red--giant stars with initial masses above 9 M$_{\odot}$ 
(Li et al. 2011a). 

The case of SN 2011fe arose great expectations, since the SN was
much closer than in previous occasions. There was, however, no object
seen at the location of the supernova in pre--explosion images,
down to magnitude 27.4 (in the ACS/F435W band) (Li et al. 2011a). 
The conclusion of the analysis is that, for SN 2011fe, the red giant
progenitor is excluded, while  a subgiant or a main--sequence companion star 
still are possible progenitors, from the imaging approach.  

Very recently, the nearby supernova SN 2014J in M82 has been tested in the same
way. This supernova is only at 3.5 Mpc. Elias--Rosa, Greggio \& Botticella  
(2014) have analyzed deep archival {\it HST} WFC3/IR images 
of M82 in the F110W and F160W filters, taken in Jan. 2010, and used them in an 
attempt to identify a progenitor for the SN, by registering the {\it HST} 
images 
with images of the SN taken on Jan. 23 2014. As in the SN 2011fe case,  
they can again exclude a red--giant companion. The limits are 
consistent with the companion being (if not another WD) a subgiant or a
main--sequence star. Goobar et al (2014) have also explored the {\it HST} 
images of the  explosion region. The observational limits, however are not as 
constraining here as in the case of SN 2011fe.

\section{Seeing the companion through the early light curve}

\begin{figure}
\centering
\includegraphics[width=0.6\columnwidth]{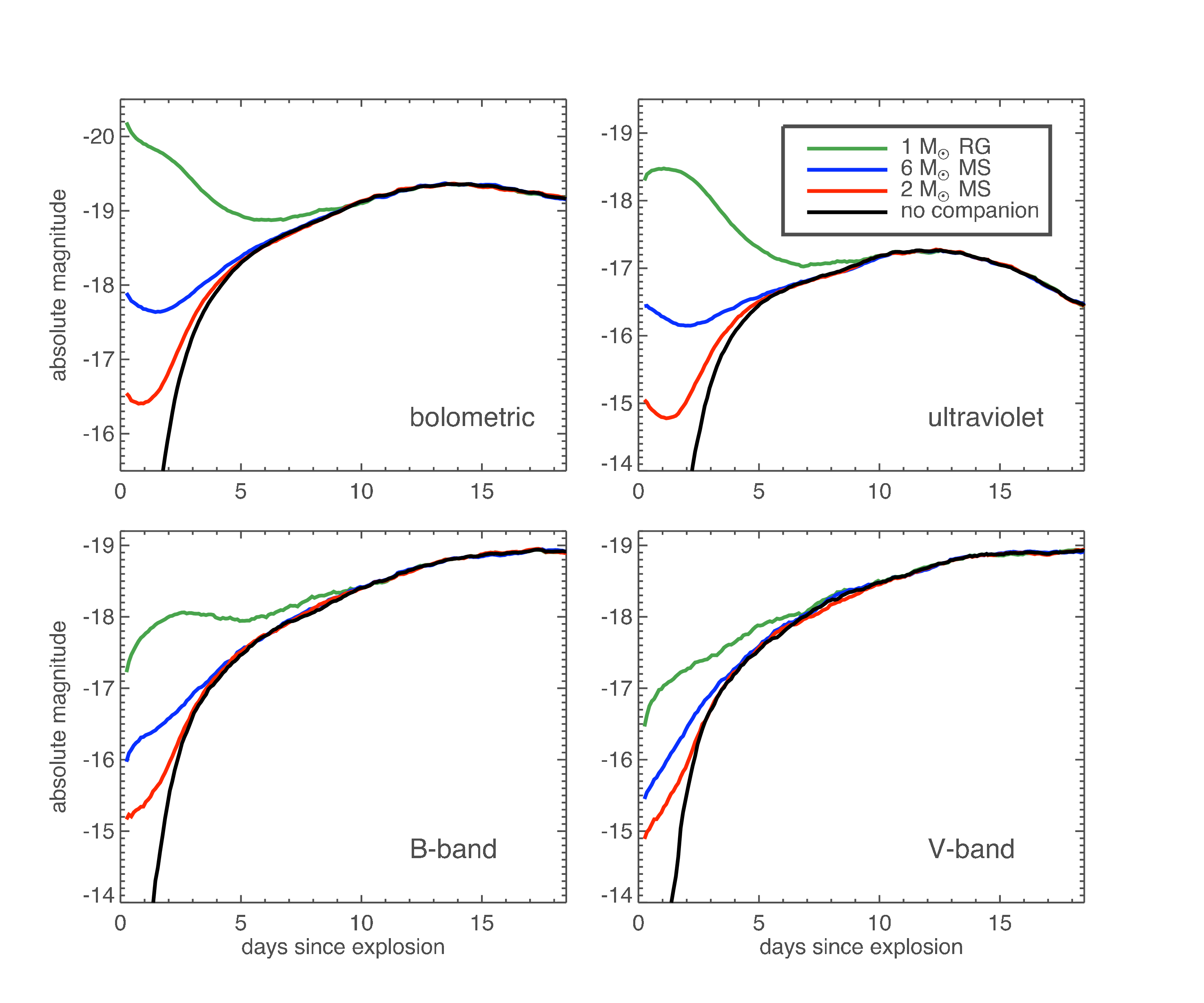}
\caption{Differences in the early light curve of a SN Ia, depending 
on the nature of the companion star, from Kasen (2010). Four possible 
progenitor scenarios are modelled: a RG companion at a distance $a = 2 
\times 10^{13}$ cm (green lines); a 6 M$_{\odot}$ MS companion at 
$ a = 2 \times  10^{12}$ cm (blue lines); a 2 M$_{\odot}$ MS companion at 
$a = 5 \times 10^{11}$ cm (red lines), and the lack of companion 
(black lines). (Courtesy of Dan Kasen. \copyright AAS. Reproduced with 
permission).}

\label{Figure 4}
\end{figure}

According to Kasen (2010), the impact of the supernova debris on 
the companion produces a bright X--ray (0.1--2 keV) burst lasting from minutes 
to hours. The diffusion of this X--ray emission gives rise to a longer--lasting 
optical/UV emisson which exceeds the radioactively powered emission from the
supernova for the first few days after the explosion. This effect can be seen 
in Figure 4. The signatures are prominent for viewing angles 
looking down upon the shocked region, which should be about a 10\% of the 
times.  Kasen (2010) concludes that the current optical and UV data do 
effectively constrain the red giant companion channel, disfavoring it. 

Zheng et al. (2014) (see their Figure 3), have presented very early 
light--curve data from SN 2014J. When comparing them with the predictions of 
Kasen (2010), the non--companion case appears favored. 

High--z searches might provide the tightest constraint on the SN progenitors. 
Goldhaber et al. (2001) show, in their Figure 1, the light curves of 35 
high--z SNe Ia found by the {\it Supernova Cosmology Project (SCP)}. These 
data, as well as  more recent ones (Conley et al. 2006; 
Hayden et al. 2010; Bianco et al. 2012) do not show any evidence for a 
companion in the early light curves of high--z SNe Ia samples. Hayden et
al. (2010), from a simulation of the shock interaction with a companion, rule
stars with masses larger that 6 M$_{\odot}$ and also disfavour red--giant
companions.

\section{Search for companion stars of SNe Ia in our Galaxy and in the LMC}

Another method to identify the progenitors of the SNe Ia was proposed by 
Ruiz--Lapuente (1997): to inspect the stars within the innermost regions of 
the Galactic SNe Ia remnants in search of the mass--donor star (to either 
find it or to show its absence), in the area where it should still remain 
after the explosion, moving with a peculiar velocity gained from the orbital 
velocity in the binary system before the explosion. A given star, to be 
candidate to donor in a SN Ia explosion, should be at the distance of the 
remnant, moving with enhanced velocity, and maybe also show signs of 
contamination by the iron--peak rich part of the supernova ejecta. A subgiant 
named Tycho G was found to be a likely candidate companion for SN 1572 
(Ruiz--Lapuente et al. 2004), since it is close to centre of the SNR, at a 
distance compatible with that of the remnant, and it is in a region where 
stars follow the rotational pattern of the Galaxy, but it has a radial 
velocity well above the 20 to 40 km s$^{-1}$, typical at the distance of 
SN 1572. It also has a high proper motion. A chemical analysis of the star 
showed enhancement of Ni in the surface (Gonz\'alez Hern\'andez et al. 2009), 
suggesting contamination by the supernova ejecta. That was disputed by 
Kerzendorf et al. (2009, 2013), who  argued that all those characteristics 
might just correspond to a chance interloper. Based on greatly improved proper 
motion measurements and a more refined chemical analysis, Bedin et al. (2014), 
however, have shown that the probability of having found such an interloper at 
random is extremely low. 

Schaefer \& Pagnotta (2012) have looked as well for a companion star in 
a SNR of the LMC (SNR 0509--67.5) and found no star that could have been the 
mass--donor in the progenitor system. Their result points to the supernova 
having resulted from merging of two white dwarfs. In addition, Edwards, 
Pagnotta \& Schaefer (2012) have examined the innermost area of the remnant 
SNR 0519--69.0, also in the LMC, and eliminated red giants, 
subgiants, and He stars as possible companions of the SN.

Gonz\'alez Hern\'andez et al. (2012) (see also Kerzendorf et al. 2012) have 
inspected the remnant of the Galactic
SN 1006, determining distances and chemical abundances for all candidate stars 
within the innermost 27\% of the area of the remnant. The lack of detection 
of any viable candidate star rules out red giant and subgiant stars, as well 
as any star brighter than M$_{V}$ $\sim$ +4.9 (approximately equal, or 
slightly less than the solar luminosity).

The key point, from the theoretical point of view, is that all groups that 
have simulated the impact of the ejecta of a supernova on its companion star 
(Marietta, Burrows \& Fryxell 2000; Pakmor et al. 2008; Pan, Ricker \& 
Taam 2012a; Liu et al. 2012, 2013) consistently find that the companion 
survives the explosion. This important conclusion is the basis for the 
observational searches.

There has been debate on whether the surviving companion of
a WD plus main sequence system, or a WD plus subgiant system,  
would show rapid rotation after the explosion (Gonz\'alez Hern\'andez et al. 
2009; Kerzendorf et al. 2009, 2013; Bedin et al. 2014). 
Hydrodynamic simulations by Pan, Ricker \& Taam  (2012a), for main sequence 
companions, show that they would lose about half of their initial angular 
momentum, their rotational velocity dropping to a quarter of the original 
rotational velocity. The simulations by Liu et al. (2012), also for a companion
on the main sequence, equally show that its rotational velocity  can be 
significantly reduced by the effects of the impact of the SN ejecta, falling 
to a 32\%--14\% of its pre--explosion value, due to remotion of 55\%--89\% 
of the initial angular momentum, taken away by the material stripped during 
the interaction with the supernova ejecta. It is easy to see (Fig. 1 of 
Marietta, Burrows \& Fryxell 2000, for instance) that, in the case of a 
1.1 M$_{\odot}$ subgiant, remotion of $\sim0.15 M_{\odot}$ by the impact of
the ejecta means reducing the radius of the star, immediately after the 
impact, to about 1/3 of its previous value only (in front of about 1/2 in 
the main sequence case), so the drop in rotational veocity must be 
correspondly larger.   

In their simulations, Pan, Ricker \& Taam (2012a) find that the contamination 
with Ni in the companion star, from the passage of the SN ejecta, is of $\sim 
10^{-5} M_{\odot}$, for a main sequence star, and of $\sim 10^{-8} M_{\odot}$ 
for a red giant. 

Another point concerns the luminosity to be expected, for the surviving 
companions of recent SNe Ia. Podsiadlowski (2003) found that, in the case 
of a subgiant, the star, 10$^{3}$--10$^{4}$ yr after the explosion, might be
either significantly overluminous or underluminous, that depending on the 
amount of heating and the amount of mass stripped, as well as on the previous
binary mass transfer. More recently, Shappee, Kochanek \& Stanek (2013) have
claimed that, in the case of a main sequence companion (and maximizing the 
heating), the object should remain significantly overluminous for the above
time lapse, but the more realistic simulations of Pan, Ricker \& Taam 
(2012b), also for a main--sequence star, predict luminosities much closer to 
that of Tycho G, $\sim500$ yr after the explosion. In the case of a subgiant, 
a larger fraction of the material should be directly stripped by the shock 
wave generated by the impact of the SN ejecta, and there should be less heating 
of the fraction of the envelope that remains bound.  

In all the preceding considerations, it has been implicitly assumed that 
there is no significant delay between the accretion phase that brings the white 
dwarf to the Chandrasekhar mass and the SN explosion. That has been questioned
by Di Stefano, Voss \& Claeys (2011), who propose a model in which the 
C+O white dwarf, spun--up by accretion of matter and angular momentum, is 
able to sustain a mass above the Chandrasekhar mass, and only reaches the 
conditions for explosive C burning when it has lost enough angular momentum, 
on a time scale that may be long enough to allow the companion star to evolve 
to the white dwarf stage. Also, overcoming the Chandrasekhar--mass limit 
would allow exhaustion of the envelope of the companion star, only its compact
core remaining at the end of the mass--transfer phase. Based on that, Di 
Stefano \& Kilic (2012) argue that the lack of evidence of ex--companion star 
in the above mentioned SNR 0509--67.5 does not mean that such companion does 
not exist, since it could have become a C+O or a He white dwarf by the time of 
the explosion. The problem of the time scale of spin down of the primary white 
dwarf has been very recently addressed by Meng \& Podsiadlowski (2013), who 
obtain an upper limit of a few 10$^{7}$ yr. Such times would still allow a 
companion star to become dimmer than the upper limit set by Schaefer \& 
Pagnotta (2012), according to Di Stefano and Kilic (2012).  

Another remnant being now studied is the Kepler SNR (SN 1604). From the lack of
bright stars in the field, Kerzendorf et al. (2014) have ruled out red 
giants as possible companions of SN 1604. VLT observations with FLAMES of the 
stars in  more than 20\% of the inner core of the SNR have now been granted 
(Ruiz--Lapuente et al. 2014). It will be very interesting to see what 
high--resolution spectra reveal. Other fairly symmetrical Galactic SNR are 
planned to undergo a similar scrutiny

\section{SN Ia candidates from orbit reconstruction}

Another approach to dilucidate the progenitors of SNe Ia is to reconstruct 
the orbit of systems containing a WD. Along these lines, the tight binary 
system CD--30$^{0}$ 11223 has been found to consist of a C+O white dwarf 
plus  a hot helium star (Geier et al. 2013). The system turns out to be a 
progenitor candidate for the double detonation SN Ia scenario (see Section 
2.2). Wang \& Han (2012) have studied this kind of possible progenitor 
system (see also Wang et al. 2009). A C+O white dwarf is first formed, from 
the initially more massive star, and the system then is in a close orbit. The 
mass donor later reaches Roche--lobe overflow and becomes a He star, but 
evolves, after exhaustion of the central He, to the red giant stage. The 
system thus becomes a C+O WD plus a He red giant. The mechanism can also work 
with the He star still staying in the main sequence phase. Fink, Hillebrandt 
\& R\"opke (2007) reproduce, in a 3D simulation of a double detonation, these 
explosions. They find that the He detonation in a shell succesfully  gives 
rise to a second detonation in the C+O core. In the outcome, $^{56}$Ni masses
about 0.40--0.45 M$_{\odot}$ are produced, with rapidly expanding  
$^{56}$Ni in the outer layers.  They note, however, the lack of observations of 
this type of explosion (SN 1991T could resemble it, but the core 
contained 0.8 M$_{\odot}$ of $^{56}$Ni). 

A C+O WD plus a He donor could, instead, be the progenitor of the 
so--called ``Type .Ia'' supernovae. We do not include these in our physical 
diagram for SNe Ia, however, since their likely He features at maximum (Kasiwal 
et al. 2010, and references therein) rule them out as possible SNe Ia (we 
only consider, in this paper, the types of explosions, either total or 
partial, that by their features at maximum can be regarded to be such).

\section{Outliers from the brightness--decline rate relation and the bulk 
of SNe Ia}

From the many systematic searches made at various redshifts, it 
has been possible to identify SNe Ia that fall well away from the Phillips 
(1993, 1999) relationship between the peak luminosiy and the rate of 
decline of the light curve. Such relationship was traditionally related 
to the amount of $^{56}$Ni synthesized in Chandrasekhar--mass models, since its
variation not only correlated with that of the maximum luminosity, but it also 
produced opacity variation in the envelope of the SN, which resulted in 
slower decline rates of the light curve for larger Ni masses. In such view 
(H\"oflich \& Khokhlov 1996; Pinto \& Eastman 2000; Bravo et al. 2009), only 
variations among Chandrasekhar--mass WD explosions  were the cause
of the relationship.  A new proposal, which completely changes the 
explanation, is to assume that the Phillips relationship results from 
variation in the viewing angle of the family of detonations of merging 
sub--Chandrasekhar explosions (Moll et al. 2013). That marks a new turn in the 
search for the physical basis of a relationship that is crucial for 
cosmology, and it thus calls for further investigation. Since this is relevant
for cosmology, we mention that the method of 
determination of H$_{0}$ using nebular spectra of SNe Ia (Ruiz--Lapuente 1996) 
favored a value of 68 km $\pm$ 7 (stat) $\pm$ 1 (updated systematic error)
km  s$^{-1}$ Mpc$^{-1}$, in good agreement with the 
latest results from the {\it Planck} satellite (Ade et al. 2013). The  light 
curves of SNe Ia (H\"oflich \& Khokhlov 1996) favored 67 $\pm$ 9 km s$^{-1}$ 
Mpc$^{-1}$ (also in agreement with {\it Planck}). Riess et al. (2011), within
the {\it SH0ES} program, had found $H_{0} = 73.8 \pm 2.4$ km s$^{-1}$ Mpc$^{-1}$,
in tension with the {\it Planck} result (but see, more recently, Riees 2014). 

\subsection{Super-Chandrasekhar SNe Ia}

A few discoveries of highly luminous ($M_{V} \sim$ -20.4) SNe Ia suggest 
the existence of super--Chandrasekhar mass explosions (Howell et al. 2006).
These supernovae show very slowly evolving Si II $\lambda$ 6355 \AA \ 
absorption velocity, and they can also show a plateau in their blue light 
curve (Scalzo et. al. 2012). The very large mass of $^{56}$Ni  needed to 
explain those events can plausibly be produced by the collision of two white 
dwarfs (Raskin et al. 2010) or by accretion on a rapidly rotating C+O WD. The 
{\it SN Factory} has tried to evaluate the percentage of super--Chandrasekhar 
SNe Ia, and they suggest about 2\% (Aldering 2011). A super--Chandraskhar WD 
can be formed if it is supported by rapid rotation (Hachisu 1986; see also 
Section 10), and the rotating WD is more massive than a non--rotating WD with 
the same central density (Yoon \& Langer 2005). For a given  central density, 
the density profile is shallower for those more massive WDs, and therefore the 
mass contained within the density range for $^{56}$Ni production is larger. 
Also, for the same central density, a flame produces more $^{56}$Ni due to 
less pre--expansion ahead of the propagating flame. In the case of merging, 
one can have tamped detonations (Howell et al. 2006). Those tamped detonations 
in rapidly rotating WDs can synthesize amounts of $^{56}$Ni as high as 1.6--2
M$_{\odot}$. Taubenberger et al. (2011) have estimated that the total mass 
of the WD, in the case of SN 2009dc, was $\sim2.8 M_{\odot}$, and the ejected 
$^{56}$Ni mass was $\sim1.8 M_{\odot}$.
 
The fact that some of those events show C II absorption features in their 
spectra reinforces the hypothesis that they come from mergings of two WDs.

On the other hand, from modeling (Hillebrandt et al. 2013) of the violent WD 
merger scenario (Pakmor et al. 2010), it seems unlikely that these supernovae 
would come from violent double--degenerate mergers. The $^{56}$Ni mass 
produced only depends on the mass of the primary WD, there. Since exploding 
C+O WDs, in the violent merger model, usually have masses well below 1.3 
M$_{\odot}$ (Ruiter et al. 2013), that limits the amount of $^{56}$Ni produced 
in the explosion to $\sim1 M_{\odot}$ only. However, recent analyses of the
nebular spectra of super--Chandrasekhar events (Taubenberger et al. 2013a), 
the case of SN 2009dc in particular, indicate that those outbursts can be 
explained by a merger of two massive C+O white dwarfs, producing 
$\sim$ 1 M$_{\odot}$ of $^{56}$Ni and $\sim$ 2 M$_{\odot}$ of ejecta. That 
would come from the explosion of a Chandrasekhar--mass white dwarf, enshrouded
by 0.6--0.7 M$_{\odot}$ of C+O--rich material.

\subsection{Supernovae strongly interacting with the CSM}

\begin{figure}
\centering
\includegraphics[width=0.6\columnwidth]{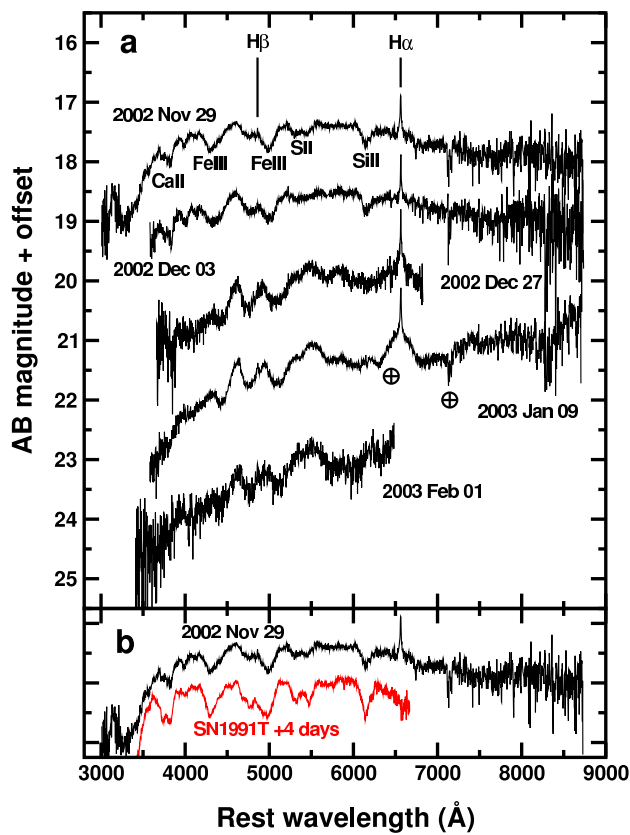}
\caption{Spectroscopic evolution of SN 2002ic (see Hamuy et al. 2003).
The spectra were taken at Las Campanas Observatory. In panel a, the evolution 
for +6, +10, +34, +47, and +70 days from estimated maximum light is shown. The 
top spectrum displays the Si II $\lambda$ 6355 \AA\ feature typical of the 
SNe Ia class, as well as prominent Fe III absorption features at 
$\lambda\lambda$ 4200 and 4900 \AA. In panel b, a comparison of the spectrum 
of SN 2002ic at the epoch +6 days from maximum with a spectrum of
SN 1991T, obtained at the +4 days epoch, shows similarity, except for the 
H$\alpha$ emission, which is not present in SN 1991T.
(Courtesy of Mario Hamuy. \copyright Nature Publishing Group. 
Reproduced with permission).} 

\label{Figure 5}
\end{figure}

There is a fraction of supernovae which show  narrow hydrogen 
emission lines. They were first noticed by Hamuy et al. (2003), in
SN 2002ic (see Figure 3). Such SN have been labelled in various ways, until 
recently being dubbed SNe Ia--CMS (Silverman et al. 2013). The existence of 
the SN Ia--CSM class of objects seems to indicate that at least some SNe Ia 
do arise 
from the SD channel, since a H--rich CSM can form during the evolution of SD 
systems. Hamuy et al. (2003) suggest that SN 2002ic could have arised from 
a binary system containing a C+O white dwarf plus a massive (3--7 $M_{\odot}$)
AGB star, where the total mass loss in H can reach a few solar masses, 
since their analysis of the  narrow component of H$\alpha$ implies a high 
mass--loss rate of $\sim10^{-2.4} M_{\odot} yr^{-1}$ (see Figure 5). 
The accreted mass would come from the wind of the AGB star, partially captured
by the WD. Despite the mass--loss rate being so high, however, the supernova 
has not been detected in radio.

Dilday et al (2012) find that the supernova PTF11kx is of Type Ia, and
suggest a symbiotic nova progenitor (see, however, Soker et al. 2013). Its 
late--time spectrum confirms that 
it is, indeed, a SN Ia. A time series of high--resolution spectra of this
supernova reveals  a complex circumstellar environment, with multiple 
shells similar to those ejected by Nova RS Ophiuchi. 
Dilday et al. (2012) found, from the {\it Palomar Transient Factory (PTF)}, 
that the SN Ia--CSM are about 0.1--1\% of all SNe Ia. This is more or 
less consistent with the theoretical expectations for the fraction of SNe Ia 
from the symbiotic progenitor channel: between 1 and 30\% (Han et al. 2004;
L\"u et al. 2009).

Concerning other typical characteristics, SNe Ia--CSM have peak absolute 
magnitudes in the range  --21.3 $\leq M_{R} \leq$ --19 mag, with relatively 
long rise times of $\sim20$--40 days. They do not emit neither in radio nor 
at X--ray wavelengths.  

\subsection{Types of subluminous SNe Ia}

\subsubsection{SNe Ia of the SN 1991bg--type}

SN 1991bg came as a suprise, being a subluminous SN Ia, one 
order of magnitude fainter than normal SNe Ia (Filippenko et al. 1992;
Leibundgut et al. 1993; Ruiz--Lapuente et al. 1993). The amount
of $^{56}$Ni synthesized was only about 0.07 $M_{\odot}$ (Ruiz--Lapuente 
et al. 1993). It is clearly out of the Phillips relation. Later, there 
were many more discoveries of this type, and one could start to think of a 
SN 1991bg--class. Li et al. (2011b) quantify this class as 15\% of all 
SNe Ia. 

Pakmor et al. (2011) suggest that violent mergers of WDs with a
primary of 0.9 M$_{\odot}$  reproduce very well the 1991bg--like 
SN. Indeed, the simulated optical light curves fit well the data (Hillebrandt 
et al. 2013). Very recently, the presence of 
[O I] $\lambda\lambda$6300, 6364 emission in the nebular spectrum of 
SN 2010lp, suggesting that oxygen is distributed in a non--spherical 
region close to the centre of the SN ejecta, has also been interpreted as 
the result of a violent merger (Taubenberger et al. 2013b).

\subsubsection{Ca--rich transients}

\begin{figure}
\centering
\includegraphics[width=0.6\columnwidth,angle=-90]{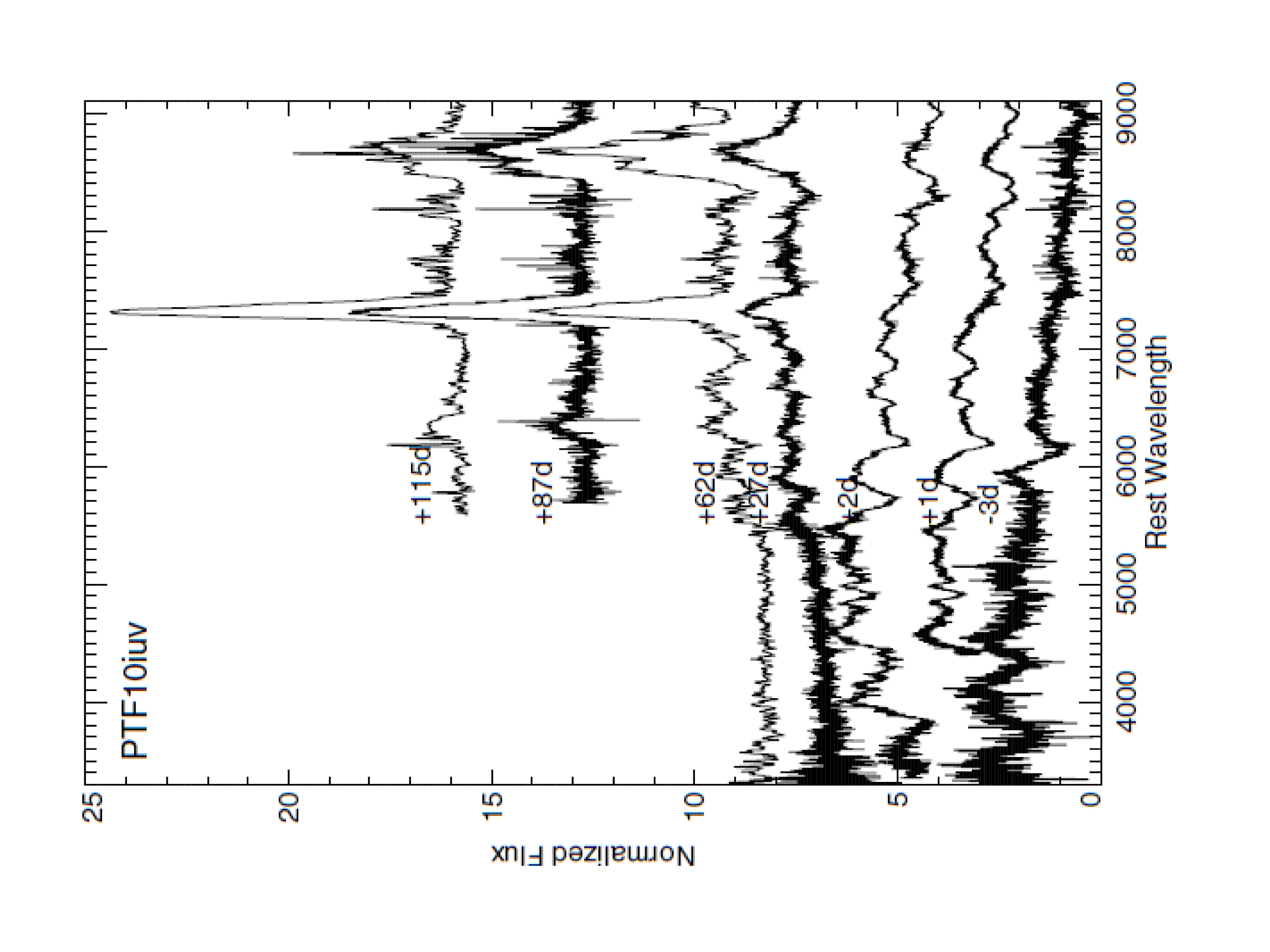}
\caption{Evolution of a Ca--rich transient. It looks like a normal SN Ia at
maximum, but it develops a Ca--dominated nebular spectrum. (Courtesy 
of Mansi Kasliwal; Kasliwal et al. 2012. \copyright AAS. Reproduced with 
permission).}
\label{Figure 6}
\end{figure}

These events do exhibit unusually strong Ca features at nebular phases,
while they look as spectroscopically normal SNe Ia at maximum. Their 
distribution within their host galaxies shows great similarity with that of 
the SNe Ia and indicate old progenitor systems (Lyman et al. 2013). According 
to these authors, they are consistent with helium--shell detonations on 
low--mass C+O  white dwarfs. The objects display low peak luminosities, 
fast photometric evolution, high ejecta velocities, strong Ca emission lines, 
and they are located in the extreme outskirts of their host galaxies (Kasliwal 
et al. 2012; see Figure 6).

\subsubsection{Type  Iax supernovae}

\begin{figure}
\centering
\includegraphics[width=0.6\columnwidth,angle=-90]{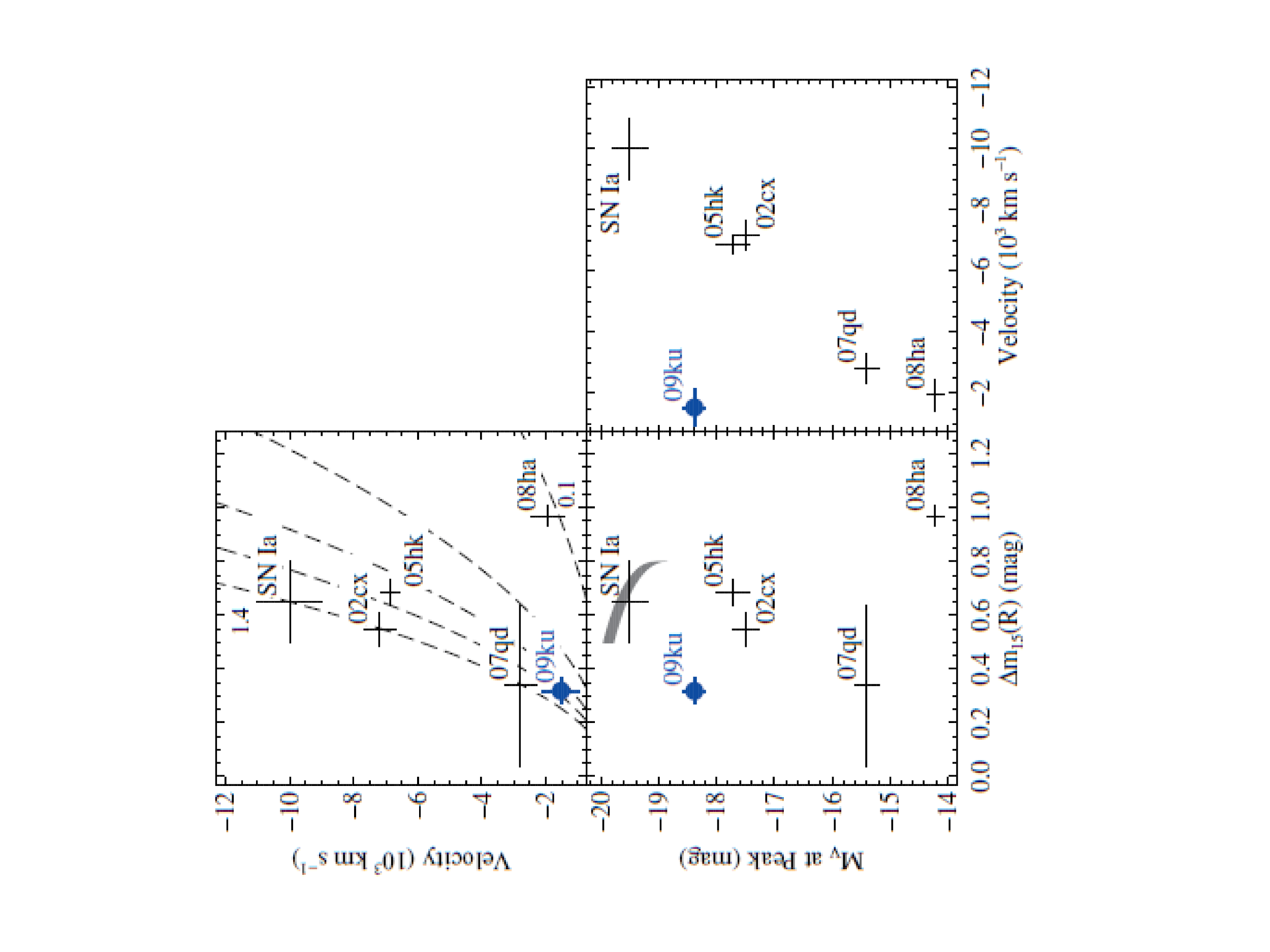}
\caption{Inhomogenety of the properties of SNIax. (Courtesy of Gautham
Narayan; Narayan et al. 2011. \copyright AAS. Reproduced with permission).}
\label{Figure 7}
\end{figure}

Foley et al. (2013) identify a new subclass of supernovae called Type Iax.
They have low maximum velocities (2000 $\lesssim v \lesssim$ 8000 km $^{-1}$),
and typically low peak magnitudes (--14.2  $\gtrsim M_{V,peak} \gtrsim$ 
--18.9 mag). In fact, this is the same family of SNe Ia identified by Li et 
al. (2003) and dubbed SN 2002cx--like SNe Ia. Foley et al. (2013) find that 
this subclass comprises 31$^{+17}_{-13}$\%. of all  SNe Ia. White et al (2014), 
however, reduce it to 5.6$^{+17}_{-3.7}$. Given the large uncertainty, in 
Figure 9 we have plotted an average of the two estimates. 

Those SN2002--cx (or SN Iax) events exhibit iron--rich spectra at early 
phases, like SN 1991T (see Ruiz--Lapuente et al. 1991 for the iron--rich 
spectra of SN 1991T), a luminosity as low as events like SN 1991bg (i.e. 2 
mag below normal Type Ia), and expansion velocities roughly half those of 
normal SNe Ia. This subclass has a small $^{56}$Ni production, as seen at late 
phases. They move fast into the nebular phase, that giving evidence of the 
small total mass ejected. A survey of the models able to produce such kind of 
explosions suggests objects made of a  0.6 M$_{\odot}$ C+O WD, with a 
layer of $\sim 0.17  M_{\odot}$ of He on top, which undergo a He detonation 
(Li et al. 2003; Wang, Justham \& Han 2013). The origin of the subclass is 
still under debate, however, because it is quite inhomogeneous (Narayan et al. 
2011). White et al. (2014) divide it in two subclasses: the 
``SN 2002--cx--like'' and the ``SN 2002-- es--like'' SNe Ia. The former tend
to appear in later--type or more irregular hosts, have more varied and 
generically dimmer luminosities, longer rise times, and they lack a Ti II 
through in their spectra, when compared to the latter.

The frequency of SN 2002cx events, as compared with normal and SN 1991bg--like 
events, has also been estimated by Perets et al. (2010). Variants of SN 2002cx 
events are seen in faint supernovae such as SN 2008ha (Foley et al. 2009, 
2010). This last 
supernova is the faintest member of its subclass. Its late--time photometry is 
consistent with the production of just a few times 10$^{-3}$ M$_{\odot}$ of 
$^{56}$Ni, similar to the estimates from the early light curve (Foley et al. 
2009). The small ejecta and $^{56}$Ni masses are consistent with a failed 
deflagration of a WD, that did not disrupt the progenitor (Jordan et al. 2012; 
Kromer et al. 2013a). There is still another model, proposed earlier, for this
particular Type Iax supernova: it is the fallback of a core--collapse 
supernova (Moriya et al. 2010). This model, then, does not treat SN 2008ha as a 
thermonuclear supernova, but as the collapse of a C+O star of 13M$_{\odot}$ 
(model 13CO2 in their paper). The boundary between the fallback region and the 
ejecta is determined by whether the velocity of the region exceeds the escape 
velocity or not. Lyman et al. (2013), argue that the host environments and 
morphologies point to a generally younger population for this subclass.
A model which synthesizes 0.003 M$_ {\odot}$ of $^{56}$Ni and ejects  
0.074 M$_{\odot}$ of material seems to reproduce the spectra and light curve 
of SN 2008ha.  Foley et al. (2013) find the environment of some of 
these Type Iax SN to be typical of old progenitor systems. They discuss that
the  most 
significant reason for their classification of SN 2008ha as a SN Ia is the 
presence of signatures of the products of thermonuclear processing of C+O, in 
particular that of sulfur lines, with an intensity that is only typical of 
SNe Ia (Foley et al. 2009). Very recently, Foley et al. (2014) report 
the possible detection of the stellar donor of SN 2008ha in images 
from the {\it HST}. Different possibilities for the progenitor remain open,
though the age is constrained to be $<$ 80 Myr. 

Thus, while Type Iax SNe, fall well below 
the Phillips relation,  some of these events 
should be regarded as subluminous SNe Ia, which might 
be linked either to He detonations in a shell or to failed deflagrations. 
Others might originate in core--collapse. 
The diversity of the events (see Figures 7, 8) suggests that several 
mechanisms take place within the sample of SNe Iax, 
explaining the properties of the different observed events.

\begin{figure}
\centering
\includegraphics[width=0.6\columnwidth]{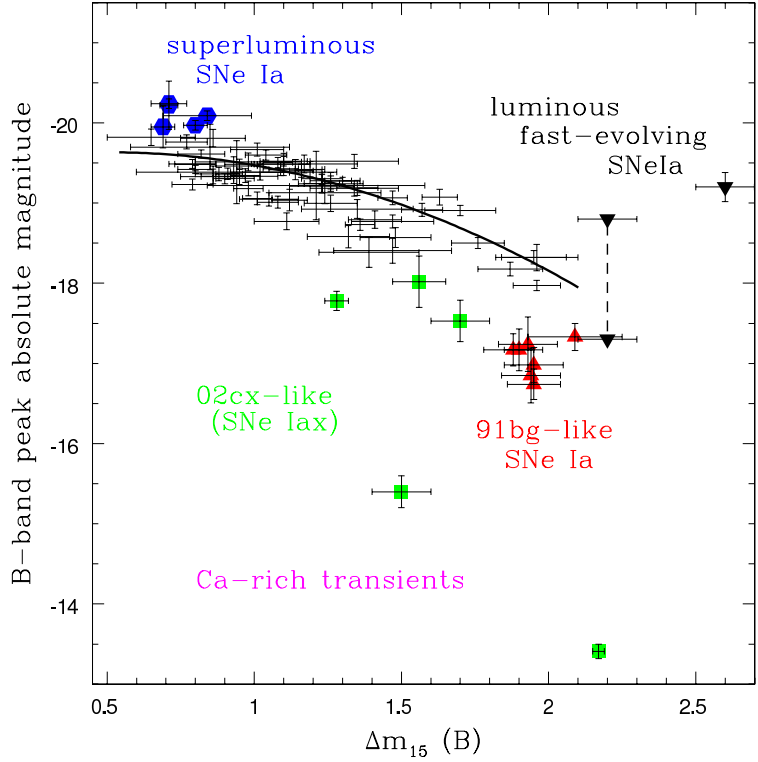}
\caption{The $M_{B}$ $vs$ $\Delta m_{15} (B)$ 
diagram using the CfA3 sample of 185 SNe Ia (Hicken et al. 2009).
The solid line is the Phillips et al. (1999) relation.
The fundamental relation developed by Phillips (1993) was
modified in Phillips et al. (1999).
The figure is adapted from Hillebrandt et al. (2013), and
we  use the same sources for superluminous and 91bg-like
subluminous events. The diversity of Type Iax is taken from 
Narayan et al. (2011). The area of the Ca--rich transients 
is marked on the diagram, following  Kasliwal  et al. (2012). 
The data on the luminous fast--evolving SNe Ia come from 
Perets et al. (2011).
The proportion in Nature of the different events is
tentatively given in Figure 9.
}

\label{Figure 8}
\end{figure}

\subsection{Rapidly declining SNe Ia}

Perets et al. (2010, 2011a,b; Poznanski et al. 2010) have identified 
a new subclass of SNe Ia: rapidly declining SNe, such as SN 1885A, 
1939C, and 2002bj, that, unlike the Iax subtype, are not faint. 
SN 2002bj had a short rise time ($<$ 7 days) and $\Delta m_{15}(B) 
\sim 3.2$, which is a post--maximum decline much faster than that of 
the bulk of SNe Ia ($\Delta m_{15} (B) < 1.7$), and even faster than 
that of SN 1991bg--like events ($1.7 < \Delta m_{15}(B) < 2$). However, 
it reached a peak magnitude $M_{B} = -18.5$, which is not faint. 

SN 1885A and 1939C arose in old environments, which points to also 
old WDs as their progenitors. In the case of SN 2002bj, the ejected 
mass appears to be low ($M_{ej} < 0.15 M_{\odot}$), that being consistent 
with the estimated mass of the SN 1885A remnant. Perets et al. (2011b) 
evaluate the frequency of the SNe in this subclass as being at least 
1--2\% of the global SNe Ia rate. Helium--shell detonations have been 
suggested as the explosion mechanism (Perets et al. 2010, 2011a; Poznanski 
et al. 2010), but its consistency with the observations still remains an 
open question. 

 \section{The bulk of SNe Ia: cosmological SNe Ia}

\begin{figure}
\centering
\includegraphics[width=0.6\columnwidth]{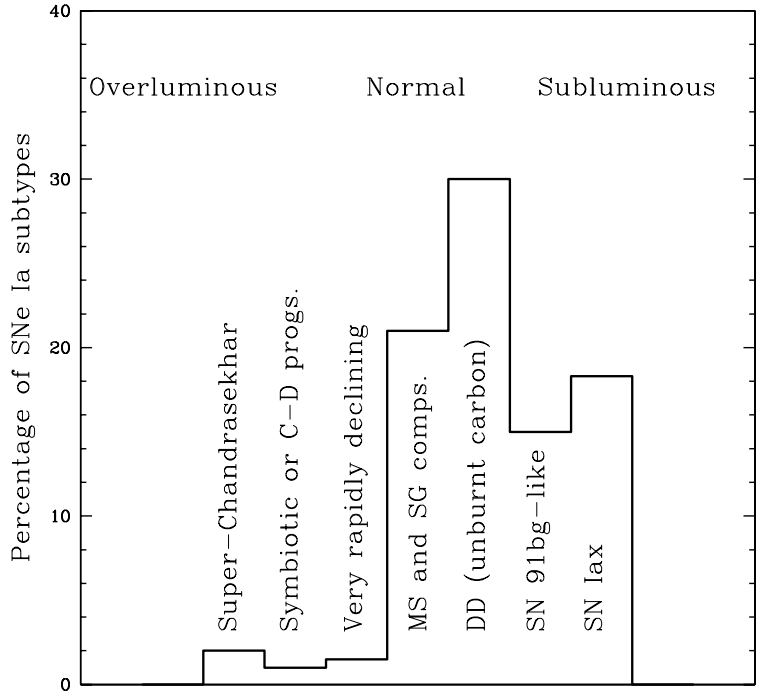}
\caption{Histogram showing the approximate percentages of SNe Ia subtypes. 
The overluminous super--Chandrasekhar SNe Ia may come from non--violent DD 
mergings. The small numbers of SN attributed either to symbiotic systems or 
to the core--degenerate scenario, those showing unburnt C in their spectra 
(likely coming from DD mergers), plus those most likely having main sequence 
stars as mass donors, together with the very fast evolving SNe Ia, have 
standard luminosities. The SN 1991bg and the SN Iax subtypes are 
underluminous. The large error bars concerning the fraction of SN of the 
SN Iax subtype (not shown in the Figure; see text) imply, of course, 
additional uncertainties on the percentages of the other subtypes, not 
affecting the relative distribution of percentages amongst them, though. 
Ca--rich transients are not included, because they are too faint to be 
considered SNe Ia}

\label{Figure 9}
\end{figure}

It has been presented in section 6 that a significant fraction (18\%)
of the population of SNe Ia show time--variable Na I D features, 
other features being variable as well. Those supernovae are normal SNe Ia, as 
the ones that we use for cosmology. Their characteristics point to recurrent 
novae (Patat 2011) or to systems with significant outflows prior to explosion,
as progenitors. Another chunk of the normal (cosmological) SNe Ia likely 
comes from double--degenerate systems ($\sim$ 30\%), as suggested by the 
unburnt C and O in the outermost layers (see section  4). 

One might ask: how can cosmological SNe Ia have two different 
origins? The reason is that in both cases the whole C+O WD is burnt, 
giving rise to the light curve whose variation in peak magnitude is
due to the amount of $^{56}$Ni synthesized in the explosion, and the
line opacity of those events modulates their rate of decline to give
$M(B)_{\rm max} = M(B)_{\rm 1.1} + 0.786[\Delta m_{\rm 15}(B) -1.1]  
 - 0.663[\Delta m_{\rm 15}(B) -1.1]^{2}$,
where $M(B)_{\rm 1.1}$ is the absolute blue magnitude, at maximum, of a 
SN Ia with $\Delta m_{\rm 15} = 1.1\ {\rm mag}$ (Phillips et al. 1999).
This relation extends the possibility of using SNe Ia with
$\Delta m_{\rm 15}$ high, since they are seen to make a continuum with
the slower ones.  
Scalzo et al. (2014) derive, from the bolometric light curves
of a sample of SNe Ia obtained by the {\it SNFactory}, a range of both 
ejected masses and $^{56}$Ni masses. Ejected masses would range, for 
normal SNe Ia, from 0.9 M$_{\odot}$ for the fastest decliners to 1.4  
M$_{\odot}$ for the slowest ones. The 91T--like SNe Ia eject masses in 
excess of 1.4  M$_{\odot}$ and produce $^{56}$Ni masses around 0.8--1.0  
M$_{\odot}$. The middle part of the sample is occupied by SNe Ia with 
Chandrasekhar--mass ejecta and producing 0.6  M$_{\odot}$ of $^{56}$Ni 
(as derived from the nebular spectra of SNe Ia in R96).

It has been argued (Dom\'{i}nguez et al. 2001) that evolutionary effects 
in the maximum--brightness--rate of decline relation could make 
difficult the use of SN Ia at very high $z$. Branch et al. (2001) propose 
an strategy to test evolutionary effects, in the prospect of a space mission 
devoted to the study of dark energy by means of high--$z$ SNe Ia. Their
proposed plan, now completed for large samples of nearby SNe Ia, is to
study all possible evolutionary effects in SNe Ia samples at low z.  
Such effects have now been evaluated from big samples, and it has been shown 
that they do not to interfere with our understanding of dark energy 
(Ruiz--Lapuente 2007). The SNe Ia at $z$ = 1.914 (Jones et al. 2013), 
at  $z$ = 1.71 (Rubin et al. 2013), and at $z$= 1.55 (Rodney et al. 2012) 
look normal and show no evolutionary effects. Concerning cosmology, it has 
been seen that these highest-z observations do suppport the $\Lambda CDM$ 
model of our Universe (see also Conley et al. 2011; Sullivan et al. 2011). 

We have learned a lot about normal (cosmological) SNe Ia from the two
recent nearby SNe Ia: SN 2011fe and SN 2014J. For the first one,
the {\it HST} deep images have allowed to discard many single--degenerate
scenarios (Li et al. 2011a,b). SN 2011fe showed unburnt C and O in the very
early spectra. One possible explanation is that it came from a 
double--degenerate progenitor (see Nomoto, Kamiya \& Nakasato 2013 for an
alternative explanation of the unburnt C and O material). For SN 2014J, 
it  has been possible to rule out red giants as the companions of
the C+O WD that exploded. 

It continues to be safe to exclude SN 1991bg--events, SNe Iax, SNe Ia-CSM
and Super--Chandrasekhar SNe Ia (Ca--rich transients as well) from the 
cosmological samples, as they lay outside the brightness--decline rate
relation used in cosmology. They might have been responsible (in particular
SNe Iax and SN 1991bg events) for some outliers present in the early samples 
(such as the first {\it SCP} SNe Ia), for which we lacked explanation
at those epochs of the cosmologically--motivated SNe Ia searches. 

Dust absorption in the host galaxy is an important source of systematic error 
in SN cosmology. A way to minimize its effect on distance determinations is to
measure the properties of the SN in the rest frame infrared. It has been 
demonstrated (Wood--Vasey et al. 2008; 
Kirshner 2010; Barone--Nugent et al. 2012) that 
SNe Ia are better standard candles in the infrared than in the optical 
wavelengths. The infrared sample of nearby SNe Ia is steadily growing, now
(Friedman et al. 2014).

\section {Conclusions}

As we have seen, the last decade has brought considerable progress in the 
still far from closed search for the progenitor systems of the SNe Ia. 
A new picture has emerged, where single--degenerate progenitors would now
make a much narrower channel than it was thought to be in the 90's. The 
highlight, here, is that the companions to SNe Ia, at the time of 
explosion, are very unlikely to be red giants or supergiants, as well as  
massive main--sequence or subgiant stars. This result comes from searches 
for companion stars in SNe Ia remnants, in our Galaxy and in the LMC; by 
looking at pre--explosion images of nearby SNe Ia, like SN 2011fe; from 
X--ray surveys of SNe Ia made with {\it Swift}; from radio observations of 
nearby SNe Ia, and from confrontation of the early light curves of SNe Ia with 
theoretical predictions.

On the other hand, the much expanded surveys for transients, at different 
wavelengths, have found from very dim explosions to overluminous outbursts. 
Multidimensional modeling of different types of thermonuclear explosions has 
reached a new level of realism, and they are confronted with observations 
uncovering the different layers of the exploding objects with unprecedented 
detail. We still lack, however, the identification of a SNe Ia with the 
dismissal of some previously observed object.

Two different channels (SD and DD) leading to SN Ia explosions still appear 
to be required, although the balance has now shifted towards the 
double--degenerate channel, where violent mergings or violent collisions of 
white dwarfs appear as a promising mechanism. The initial conditions leading 
to such collisions are still unclear, however. For subluminous SNe Ia of the 
SNe Iax type, or 
Ca--rich transients, edge--lit He detonations that might not disrupt the 
underlying C+O WD, or failed deflagrations in the outer shells, seem able to 
account for the observations and for the rates of the explosions. 
 
From all the preceding, we can reach some tentative conclusions about the 
fraction of supernovae arising from different kinds of progenitor systems 
and their explosion mechanisms (see Figure 9).  
 
$\bullet$ The supernovae that show unburnt carbon make 30\% of the
SNe Ia class. If we add to that the super--Chandrasekhar explosions (2$\%$), 
we can infer that around 32\% of thermonuclear supernovae should 
arise from the merging of two white dwarfs. Tamped detonations in rapidly
rotating white dwarfs may be the explosion mechanism, in the 
super--Chandrasekhar case, although mergings of massive white dwarfs 
could also account for these events.   

$\bullet$ Between 31$^{+17}_{-13}$ $\%$ and 5.6$^{+17}_{-3.7}$ of all SNe Ia do 
belong to the Iax subtype, likely arising either from failed deflagrations or 
from surface detonations of low--mass white dwarfs. The large error bars come 
from the uncertainty on the selection effects that work against detection of 
subluminous events. There are indications that this subclass of SNe Ia with
low ejection velocities do, in fact, split into two subclasses. Further 
research in this are is needed.   

$\bullet$ There is a 15$\%$ of SNe Ia of the 91bg--type, which are tentatively
associated with violent mergings of white dwarfs where the primary is of about 
0.9 M$_{\odot}$. Detonation of the C+O mixture should then occur.

 $\bullet$ The fraction of explosions resulting from the accretion of hydrogen 
from a red giant star seems to be very low, as metioned before. The SNe coming 
from a symbiotic progenitor should only make 0.1--1$\%$ of all SNe Ia.

$\bullet$ About 1--2\% of events have luminosities falling well in the middle 
of the bulk of SNe Ia, but they rise fast and do decline very fast after 
maximum (the fraction migh be larger, due to selection effects). Their origin
remains unclear.   

 $\bullet$ The remaining fraction of SNe Ia, a 21 $\%$, could come from  
systems made of a C+O white dwarf plus a main--sequence or subgiant 
 star, and they could 
appear, before the explosion, as supersoft X--ray sources or/and recurrent
novae. In the SN of this 
and the former group (as well as in those coming from non--violent DD 
mergings), the explosion should start as a (generally off--centre) 
deflagration, to become a detonation in reaching lower--density layers.

Given the large sample of peculiar SNe Ia, it is worth obtaining 
spectral sequences, when using  SN for cosmology. Such project would be made 
possible by dedicated space missions and by ground--based programs (Wood--Vasey 
2010).

\section{Acknowledgments}

I would like to thank Gaston Folatelli, Mario Hamuy, Dan Kasen, 
Mansi Kasliwal, Gautham Narayan, and Josh Simon, for their kind permissions
to use their Figures. My thanks also go to Hagai Perets, Lev Yungelson, 
Zhengwei Liu, Alexander Tutukov, Thomas Tauris, Enrique Garc\'{\i}a--Berro, 
Joe Lyman, Raffaella Margutti, Todd Thompson, Mickael Rigault, Markus Kromer, 
Dan Kasen, Boaz Katz, Mario Hamuy, Michael Wood--Vasey, and Marat Gilfanov, 
for valuable comments, suggestions and criticisms to the first draft of this 
paper. The useful input from an anonymous referee is also acknowledged. This 
work has been supported by Grant AYA2012--36353, from the Ministerio de 
Econom\'{\i}a y Competitividad of Spain.








\end{document}